\documentclass[lettersize,journal]{IEEEtran}

\AtBeginDocument{%
  }

\usepackage{amsmath,amsfonts}
\usepackage{algorithmic}
\usepackage{array}
\usepackage{adjustbox}
\usepackage[caption=false,font=normalsize,labelfont=sf,textfont=sf]{subfig}
\usepackage{stfloats}
\usepackage{url}
\usepackage{verbatim}
\usepackage{graphicx}
\usepackage{cite}
\usepackage{booktabs}
\usepackage{multirow}
\usepackage{colortbl}
\usepackage{microtype}
\usepackage{textcomp}
\usepackage{xcolor}
\usepackage[ruled]{algorithm2e}
\usepackage{makecell}
\usepackage{tcolorbox}
\usepackage{rotating}
\usepackage{longtable}
\usepackage{bbding}
\usepackage{tabularx}
\usepackage{pifont}
\usepackage{enumitem}
\usepackage{xspace}
\usepackage{afterpage}
\newcommand{\boxmargin}{1mm}
\newcommand{\ours}{\textsc{ACToR}\xspace}
\newcommand{\rag}{RAG\xspace}
\newcommand{\sota}{SOTA\xspace}

\newtcolorbox{myboxc}{
    colback=yellow!10!white,
    colframe=gray!50,
    arc = 0pt, outer arc = 5pt,
    boxsep=5pt, left = 3pt, right = 0pt, top = 0pt, bottom = 0pt,
    leftrule=3pt,
    bottomrule=0pt, toprule=0pt, rightrule=0pt,
    left = \boxmargin, right = \boxmargin, top = \boxmargin, bottom = \boxmargin,
    before skip=5pt,
    after skip=5pt
}

\newcommand{\mytextsuperscript}[1]{\textsuperscript{#1}}

\newcommand{\RawpromptCLSeven}{RawPrompt\textsubscript{ CodeLlama-7B}}
\newcommand{\RawragCLSeven}{RawRAG\textsubscript{ CodeLlama-7B}}
\newcommand{\RepocoderCLSeven}{RepoCoder\textsubscript{ CodeLlama-7B}}
\newcommand{\RlcoderCLSeven}{RLCoder\textsubscript{ CodeLlama-7B}}
\newcommand{\OursCLSeven}{\ours\textsubscript{ CodeLlama-7B}}

\newcommand{\RawpromptCLThirteen}{RawPrompt\textsubscript{ CodeLlama-13B}}
\newcommand{\RawragCLThirteen}{RawRAG\textsubscript{ CodeLlama-13B}}
\newcommand{\RepocoderCLThirteen}{RepoCoder\textsubscript{ CodeLlama-13B}}
\newcommand{\RlcoderCLThirteen}{RLCoder\textsubscript{ CodeLlama-13B}}
\newcommand{\OursCLThirteen}{\ours\textsubscript{ CodeLlama-13B}}

\newcommand{\RawpromptDSCOne}{RawPrompt\textsubscript{ DSCoder-1.3B}}
\newcommand{\RawragDSCOne}{RawRAG\textsubscript{ DSCoder-1.3B}}
\newcommand{\RepocoderDSCOne}{RepoCoder\textsubscript{ DSCoder-1.3B}}
\newcommand{\RlcoderDSCOne}{RLCoder\textsubscript{ DSCoder-1.3B}}
\newcommand{\OursDSCOne}{\ours\textsubscript{ DSCoder-1.3B}}

\newcommand{\RawpromptDSCSeven}{RawPrompt\textsubscript{ DSCoder-6.7B}}
\newcommand{\RawragDSCSeven}{RawRAG\textsubscript{ DSCoder-6.7B}}
\newcommand{\RepocoderDSCSeven}{RepoCoder\textsubscript{ DSCoder-6.7B}}
\newcommand{\RlcoderDSCSeven}{RLCoder\textsubscript{ DSCoder-6.7B}}
\newcommand{\OursDSCSeven}{\ours\textsubscript{ DSCoder-6.7B}}

\begin{document}

\title{Adaptive Critical Token-Aware Retrieval for Repository-Level Code Generation}
\markboth{IEEE Transactions on Software Engineering,~Vol.~XX, No.~XX, Month~20XX}%
{Duan \MakeLowercase{\textit{et al.}}: Adaptive Critical Token-Aware Retrieval for Repository-Level Code Generation}

\author{Kefeng Duan$^{\ast}$, Dewu Zheng$^{\ast}$, Yanlin Wang$^{\dagger}$, Terry Yue Zhuo, Mingwei Liu, Jianxing Yu, Jiachi Chen, Ensheng Shi, Xilin Liu, Yuchi Ma, and Zibin Zheng
\thanks{$^{\ast}$K. Duan and D. Zheng are co-first authors with equal contribution.}
\thanks{$^{\dagger}$Y. Wang is the corresponding author (e-mail: yanlin-wang@outlook.com).}
\thanks{K. Duan, D. Zheng, Y. Wang, M. Liu, and Z. Zheng are with Sun Yat-sen University, Guangzhou 510006, P.R. China (e-mail: duankf@mail2.sysu.edu.cn; zhengdw5@mail2.sysu.edu.cn; yanlin-wang@outlook.com; liumw26@mail.sysu.edu.cn; zhzibin@mail.sysu.edu.cn).}
\thanks{J. Yu is with the School of Artificial Intelligence, Sun Yat-sen University, Zhuhai 519082, P.R. China (e-mail: yujx26@mail.sysu.edu.cn).}
\thanks{T. Y. Zhuo is with Monash University and CSIRO's Data61 (e-mail: terry.zhuo@monash.edu).}
\thanks{J. Chen is with Zhejiang University (e-mail: chenjiachi@zju.edu.cn).}
\thanks{E. Shi, X. Liu, and Y. Ma are with Huawei Cloud Computing Technologies Co., Ltd. (e-mail: shiensheng@huawei.com; liuxilin3@huawei.com; mayuchi1@huawei.com).}
}

\maketitle
\begin{abstract}
The repository-level code generation task requires synthesizing code that satisfies task requirements while remaining consistent with the target repository context. Since real-world repositories often exceed the input length limits of LLMs, existing approaches commonly adopt retrieval-augmented generation (RAG) to provide repository-specific context. Despite improving repository-context retrieval, existing methods typically provide context as task-level support, without explicitly identifying the critical tokens that require fine-grained repository context during generation. During the autoregressive generation process of LLMs, errors often concentrate at a small number of decisive positions: once such tokens are generated incorrectly, subsequent code may follow an incorrect semantic path and eventually lead to functional failure. We refer to these positions as ``critical tokens''.
In this paper, we propose \ours, an adaptive critical token-aware retrieval framework for repository-level code generation. \ours identifies critical tokens during generation and triggers targeted retrieval on demand to provide repository context at these decisive positions. In addition, we design a position-aware weighting method for dense retrievers to prioritize context that is more informative for generation. We evaluate \ours on two representative repository-level benchmarks, RepoExec and CoderEval. Experimental results show that \ours consistently outperforms state-of-the-art methods, achieving relative improvements of 8.4\% on RepoExec and 15.4\% on CoderEval. Beyond performance gains, we systematically quantify the impact of critical tokens, revealing their central role in major generation failures and highlighting the necessity of targeted retrieval strategies. We provide the code and data at~\url{https://github.com/DeepSoftwareAnalytics/ACToR}.
\end{abstract}


\section{Introduction}
Recent LLMs are increasingly used in real-world software development~\cite{hou2024large,fan2023large,zheng2025towards,wang2025agents}. Unlike standalone code-generation tasks~\cite{chen2021evaluating,zhuobigcodebench}, repository-level code generation targets real-world development within existing repositories. In this setting, generated code must not only satisfy task requirements but also remain consistent with project-specific APIs, dependencies, data structures, coding conventions, and the broader repository context~\cite{yu2024codereval,xie2025repost,wang2025rlcoder}.

Therefore, repository context is essential for resolving project-specific dependencies and preserving consistency with the existing codebase. However, real-world code repositories often exceed the input length limits of LLMs, making it infeasible to provide the entire repository as input. To address this challenge, existing studies have widely adopted retrieval-augmented generation (RAG), which first retrieves task-relevant context from the repository and then incorporates it into the generation prompt. Representative repository-level RAG methods have improved contextual support from several perspectives, including context quality and retrieval decision-making. To improve context quality, RepoCoder~\cite{zhang2023repocoder} iteratively constructs new retrieval queries using generated results from previous rounds, Cocomic~\cite{ding2024cocomic} locates relevant code dependencies based on a project context graph, and DRACO~\cite{cheng2024dataflow} retrieves fine-grained code entities through dataflow analysis. In retrieval decision-making, RepoFormer~\cite{wu2024repoformer} determines whether retrieval should be triggered for the current task, RLPG~\cite{shrivastava2023repository} identifies the type of dependency context required by the task, and ProCC~\cite{tan2024prompt} dynamically selects suitable retrieval prompt construction strategies. Overall, these methods improve repository-level code generation mainly by enhancing how repository context is obtained, selected, and organized.

Although the above methods improve the quality of repository context from the perspectives of retrieval strategy, context source, and retrieval decision-making, they still organize and use context mainly at the task level, lacking explicit modeling of key token-level decisions during generation. Specifically, existing methods typically treat the retrieved context as holistic support for the entire generation task, implicitly assuming that a set of relevant repository contexts can continuously serve the subsequent generation process. However, code generation is a highly path-dependent autoregressive process, and different generation positions may require different contextual information. 

In repository-level code generation, an incorrect API name, variable reference, index, or control-flow token can cause subsequent generation to deviate from repository semantics and eventually fail the entire implementation. In other words, code generation errors are often not accumulated uniformly across all tokens, but instead concentrate at a small number of positions that have a decisive impact on the final result. We call these positions \textit{critical tokens}: \textbf{a small number of tokens in the LLM inference chain whose correctness plays a decisive role in the correctness of the final answer, and whose incorrect generation may cause the entire generated result to fail.} Therefore, the core limitation of existing retrieval pipelines is not merely that retrieval occurs before generation or remains at the query level, but that they lack a mechanism to identify these critical tokens and provide targeted repository context support at these key positions.

Based on this observation, we introduce \ours (\textbf{A}daptive \textbf{C}ritical \textbf{To}ken-aware \textbf{R}etrieval), a dynamic, fine-grained retrieval framework. Unlike static methods, our framework operates during the generation process. It identifies critical tokens where the model requires additional context and performs targeted, on-demand retrieval for each one. This adaptive mechanism ensures that the model always has the precise information it needs, thereby mitigating the quality degradation and functional failures common in existing systems. Besides, we propose a position-aware weighting method for dense retrievers. This allows us to retrieve context that is more informative for generation.

We validate our approach on two standard repository-level code generation benchmarks: RepoExec~\cite{le2025impacts} and CoderEval~\cite{yu2024codereval}.
\ours{} consistently improves over strong repository-level baselines, with relative gains of 8.4\% on RepoExec and 15.4\% on CoderEval.
Complementing these end-to-end results, we provide a quantitative characterization of \emph{critical tokens}: they occupy only a small fraction of generated positions (about 5--11\% across models), yet their composition and syntactic profiles concentrate much of the model's error and uncertainty, motivating sparse, generation-time retrieval updates rather than repository context supplied only once up front.
Our analysis further ties major failures to this small set of pivotal positions, underscoring the necessity of targeted retrieval strategies like \ours.

We summarize our contributions as follows:
\begin{itemize}[leftmargin=10pt]
    \item We introduce the concept of critical tokens for repository-level code generation. To our knowledge, this is the first work to systematically quantify the decisive impact these tokens have on the quality of the final generated code.
    
    \item We propose a novel targeted retrieval framework designed for these critical tokens. This framework performs on-demand retrieval to supply the language model with precise, timely context at the most crucial steps of the generation process.

    \item We design a position-aware weighted pooling method to enhance our retrieval mechanism. This technique refines the context embedding process, significantly boosting the precision of retrieval.

    \item We conduct extensive experiments on representative repository-level code generation benchmarks, demonstrating that our approach significantly and consistently outperforms existing SOTA methods.

\end{itemize}

\section{Motivating Examples}

This section uses two real examples from CoderEval~\cite{yu2024codereval} to illustrate two key properties of critical tokens. First, although any token change may affect subsequent generation, critical tokens are distinctive because their errors often change the model's core judgment about the current implementation logic, causing later code to follow an incorrect semantic path and eventually leading to function-level failure. Second, critical tokens are not limited to a fixed syntactic form; they may appear as API names, variable references, index expressions, or even semantic cues in comments. Therefore, a single pre-generation retrieval step or a fixed syntactic rule can hardly cover the tokens that truly determine the final generation result.

\begin{figure}[!t]
    \centering
    \includegraphics[width=0.5\textwidth]{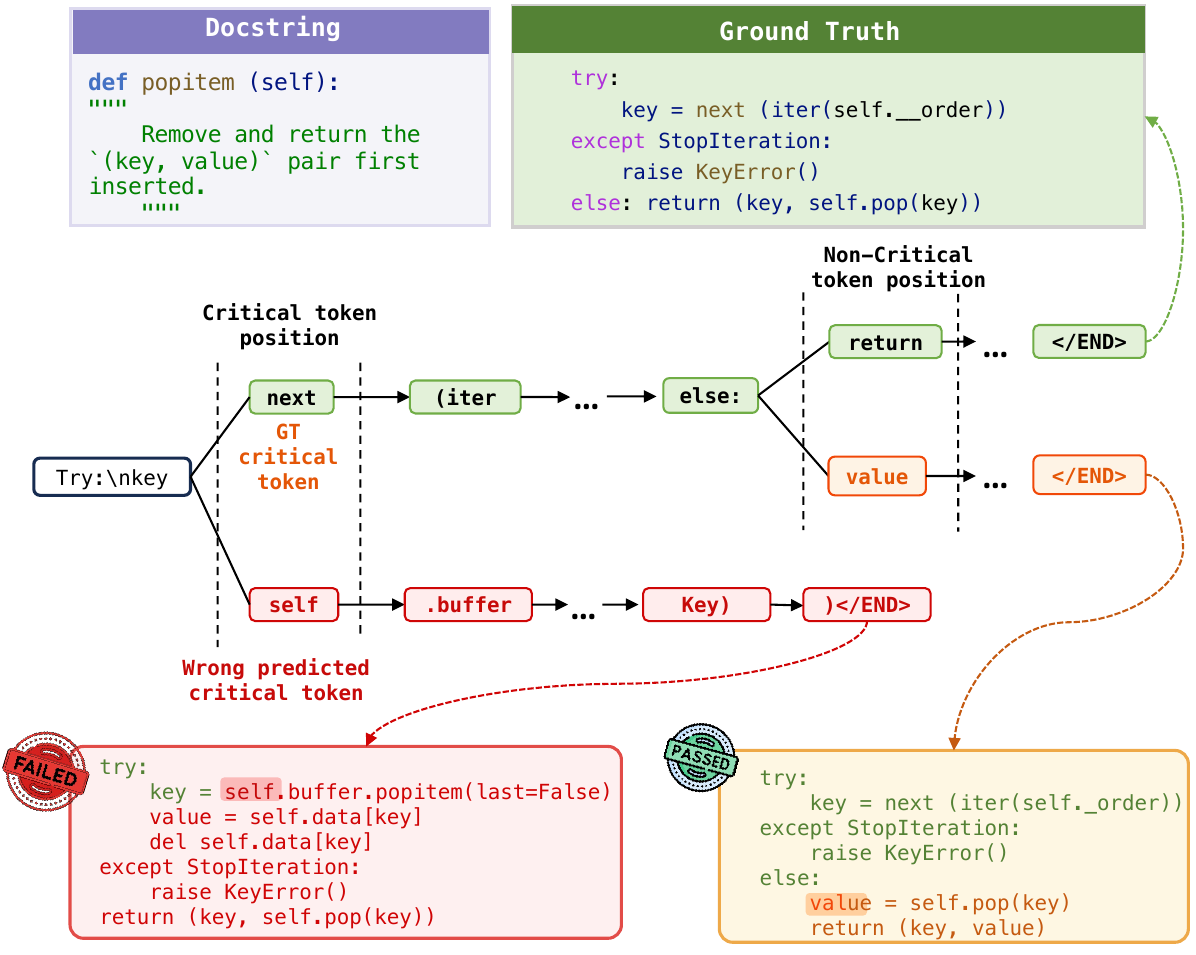}
    \vspace{-5pt}
    \caption{Motivating example from CoderEval task \texttt{62b8d22f}~\cite{yu2024codereval}: an incorrect critical token can cause catastrophic failure in subsequent generations (\textit{popitem} scenario).}
    \label{fig:example1}
\end{figure}

\begin{figure}[!t]
    \centering
    \includegraphics[width=0.5\textwidth]{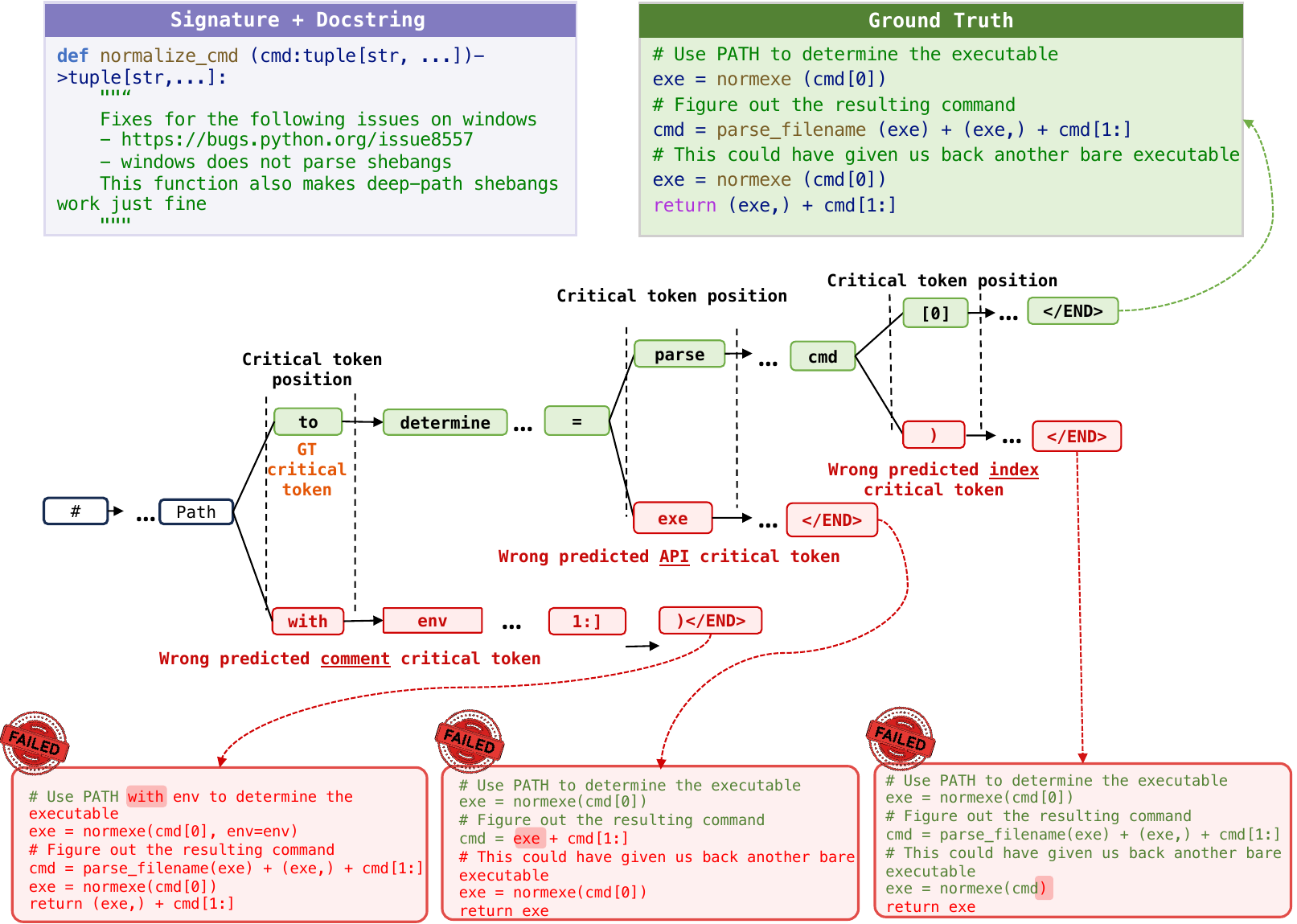}
    \vspace{-5pt}
    \caption{Motivating example from CoderEval task \texttt{62e4fb4d}~\cite{yu2024codereval}: critical tokens exhibit diverse syntax and can appear in multiple roles (\textit{normalize\_cmd} scenario).}
    \label{fig:example2}
\end{figure}

\textbf{Motivating example 1:} Figure~\ref{fig:example1} shows the difference between a critical-token error and an ordinary token error in terms of their consequences. The goal of the \textit{popitem} function is to remove and return the earliest inserted item from a cache. This behavior depends on a repository-specific ordering structure: the correct implementation should first obtain the earliest key from \texttt{self.\_order}, for example through \texttt{next(iter(self.\_order))}, and then update the cache according to that key. However, at the highlighted position, the model generates \texttt{self} instead of \texttt{next}. This error directly changes the model's judgment about which repository object the subsequent code should be built around. Instead of constructing an iterator over \texttt{self.\_order}, the model turns to an incorrect call such as \texttt{self.buffer.popitem...}. As a result, the generated function no longer follows the repository's insertion-order deletion logic; the remaining code is generated around this wrong operation and eventually deviates completely from the intended behavior. In contrast, the non-critical token error in the figure affects only a local type-related detail and does not break the main control-flow or data-flow structure of the function. This comparison shows that critical tokens are not simply tokens predicted incorrectly; they are positions where an error can make the model depart from the core implementation logic and thereby compromise functional correctness.

\textbf{Motivating example 2:} Figure~\ref{fig:example2} further shows the syntactic and semantic diversity of critical tokens. The \textit{normalize\_cmd} function needs to normalize a command representation according to the repository's existing implementation, including how executable names are parsed, how command tuples are indexed, and which internal helper APIs should be invoked. Therefore, the task requires not only understanding the function description but also following the conventions encoded in the surrounding code. In the comment-related case, an incorrect comment-related token misleads the model into inferring that an additional environment argument is needed, causing it to hallucinate an \texttt{env} parameter that is not supported by the repository API. In the API-related case, the model fails to generate the correct call to \texttt{parse\_filename} and instead falls back to simple string concatenation, such as \texttt{cmd = exe + cmd[1:]}. This bypasses the repository-defined filename parsing logic and makes the normalization process inconsistent with the project implementation. In the index-related case, the model incorrectly uses the entire \texttt{cmd} object instead of accessing a specific element, for example passing \texttt{cmd} directly to \texttt{normexe}, which expects a single element. This indexing error causes a type mismatch, propagates to the return statement, and eventually leads to a runtime failure. Although these three errors involve comments, API calls, and index expressions, they share the same property: each occurs at a position that determines the direction of the subsequent implementation and causes the generated result to deviate from repository semantics.

Together, these examples show that critical tokens have two important properties. On the one hand, \textbf{their errors can make the model continue generation along an incorrect semantic path, causing function-level failures that go far beyond a local token error}. On the other hand, \textbf{they are syntactically diverse and cannot be fully captured by a fixed syntactic category or a single pre-generation retrieval step.} These observations motivate a generation-time mechanism that identifies critical tokens as they arise and provides targeted repository context support at those key positions.

\section{Methodology}

\subsection{Overall Pipeline}

In this section, we introduce an innovative methodology for repository-level code generation, centered on a targeted retrieval mechanism guided by critical tokens. The methodology is divided into two phases: offline training and online inference. In the offline phase, we focus on training a lightweight discriminative ensemble model. Its purpose is to accurately identify critical tokens. During the online inference phase, the methodology enhances code quality through two key techniques: Position-Aware Weighted Retriever and Critical Token-Guided Inference. Figure~\ref{fig:pip} shows the overall pipeline of our study.

\subsection{Offline Training}

\begin{figure*}[!t]
    \centering
    \includegraphics[width=0.9\textwidth]{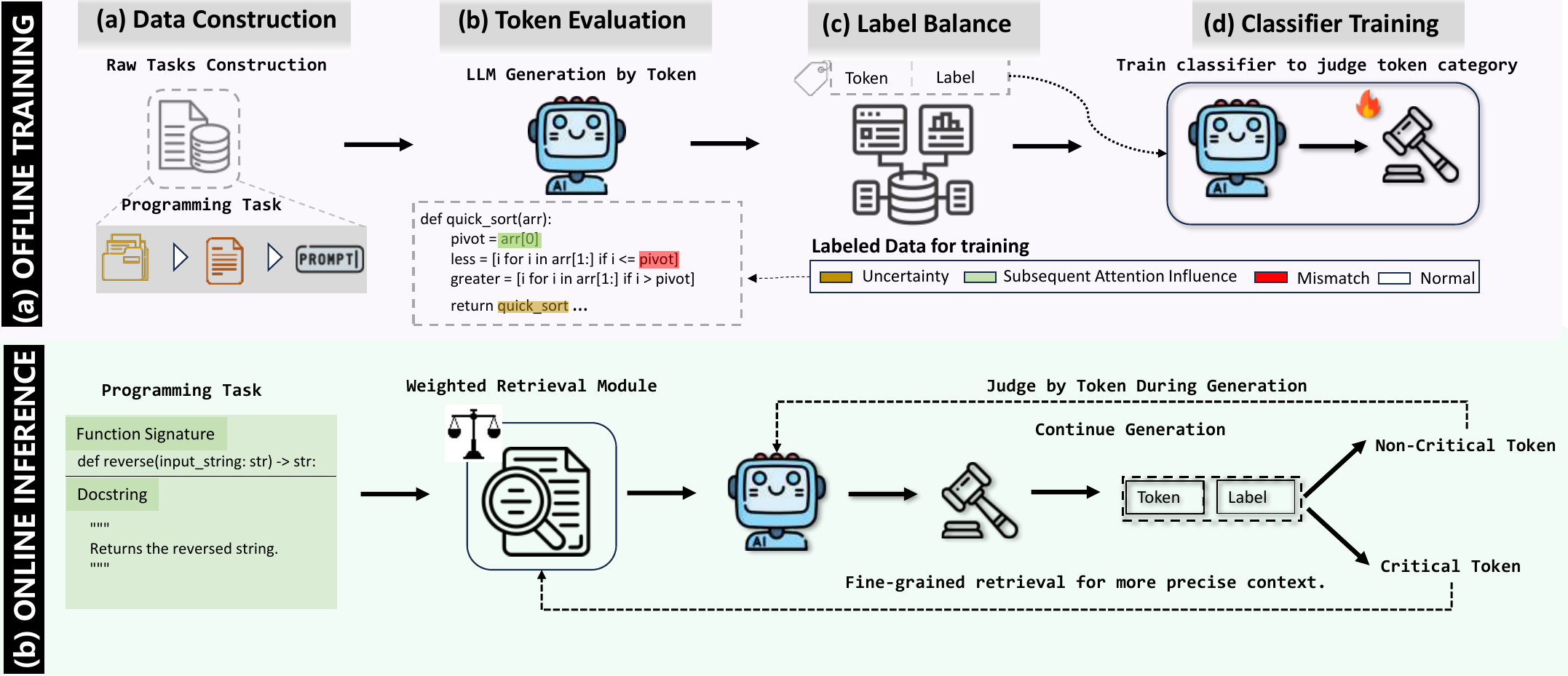}
    \vspace{-5pt}
    \caption{The Pipeline of \ours.}
    \label{fig:pip}
\end{figure*}

During the training phase, we design a critical token discriminator that identifies critical tokens and triggers dynamic retrieval based on the model’s hidden states during generation. To obtain high-quality training data that match this scenario, we follow the data construction pipeline described below (see Figure~\ref{fig:pip}).

\subsubsection{Data Construction}

This subsection describes the procedure for building our training set, which consists of three stages: Repository Filtering, Function Sampling, and Prompt Construction.

\begin{enumerate}[leftmargin=15pt]
    \item \textbf{Repository Filtering.} We select our data from RepoST-Train~\cite{xie2025repost}, a large-scale training dataset specialized for repository-level code generation, containing 7,415 functions from 824 repositories. To ensure data quality, we filter repositories according to the following criteria:

    \textcolor{green}{\ding{51}} The repository must contain more than fifteen training samples (functions).
    
    \textcolor{green}{\ding{51}} The repository must have more than ten stars on GitHub.

    \textcolor{green}{\ding{51}} The repository must be a real-world software project, rather than a collection of solutions for algorithmic challenges (e.g., LeetCode) or academic assignments.
    
    \textcolor{red}{\ding{55}} The repository must not appear in our evaluation benchmarks (RepoExec~\cite{le2025impacts} and CoderEval~\cite{yu2024codereval}).

    After applying these filters, we include 102 repositories for constructing training data.
    
    \item \textbf{Function Sampling.} After repository filtering, we further sample candidate functions within the selected repositories according to the following rules:

    \textcolor{green}{\ding{51}} The sample must include a function signature.
    
    \textcolor{green}{\ding{51}} The sample must include a complete docstring (used to describe function behavior and constraints).

   Our training dataset excludes samples that do not meet both conditions. After that, we select 1,203 task samples for our training.

    \item \textbf{Prompt Construction.} For each repository,  we follow RLCoder~\cite{wang2025rlcoder} and adopt the ``Split and Aggregate candidate'' strategy to construct repository context candidates. For each task sample, we concatenate the function signature and docstring to form the initial code sequence $X$. This sequence $X$ serves as the query to a dense retriever. The retriever ranks these candidates by cosine similarity and returns an ordered set of context snippets: $\mathcal{C}=\{C_1, C_2, \dots, C_m\}$, which is sorted by descending similarity. We then concatenate the contexts and the initial code sequence in retrieval score order to form prompts as: $P = [C_1 \;|\; C_2 \;|\; \dots \;|\; C_m \;|\; X]$, where ``$|$'' denotes a separator between contexts. In practice, we use file-path annotations of each context and the target function prompt as explicit separators to preserve context boundaries and improve readability. 
    
    After that, we treat the corresponding function body as the ground truth and record both the prompt and ground truth for training.
    
\end{enumerate}

\subsubsection{Token Evaluation}
\label{sec:token_eval}

At this stage, our goal is to determine, for each generated token, whether it is a critical token and to construct a labeled dataset for this classification task. A critical token is a token that the model tends to mispredict and whose erroneous generation would significantly affect subsequent tokens, substantially increasing the risk of functional failure in the generated code. In other words, an error at such a position is not only incorrect locally but also likely to trigger cascading errors or to break program semantics. Based on this definition, we quantify each generated token using the following three criteria:

\begin{itemize}[leftmargin=10pt]
    \item \textbf{Token Mismatch.} The model’s top-1 prediction at position $i$ differs from the ground-truth token. The formula can be expressed as:
    \begin{equation}
        x_i\ne\hat{x}_i
        \label{eq:token_mismatch}
    \end{equation}
    where $x_i$ is the ground-truth token at position $i$ and $\hat{x}_i$ denotes the model’s top-1 prediction. A top-1 mismatch indicates high prediction difficulty at this position. If a downstream impact accompanies this mismatch, the token is more likely to be critical.
    \item \textbf{Uncertainty.} We introduce entropy to quantify the model's uncertainty when generating the token at position $i$:
    \begin{equation}
        \mathcal{H}_i \;=\; -\sum_{v\in\mathcal{V}} p_i(v)\log p_i(v),
    \end{equation}
    where \(p_i(v)\) denotes the probability assigned by the model to token $v$ at position $i$, and $\mathcal{V}$ is the vocabulary of large language models. A larger entropy \(\mathcal{H}_i\) indicates greater model uncertainty at that position, and therefore a higher likelihood that the token is critical.

    \item \textbf{Subsequent Attention Influence.} We propose a metric based on the model's self-attention mechanism~\cite{vaswani2017attention} to assess the potential influence of a token on subsequent tokens.     Self-attention, the core mechanism of Transformer-based LLMs, is computed as:
    \begin{equation}
        A=\mathrm{softmax}\left(\mathrm{mask}\left(\frac{QK^\top}{\sqrt{d_k}}\right)\right)
        \label{eq:attention_metrix}
    \end{equation}
    where $A_{j,i}$ denotes the attention weight from position $j$ (Query) to position $i$ (Key/Value). Considering up to the next $k$ tokens (or until the sequence end), we define the subsequent attention on position $i$ as:
    \begin{equation}
        a_{\mathrm{mean}}(i)=\mathrm{Mean}_{j\in\{i+1,\ldots,i+k\}}Top_{K}(A_{j,i}).
        \label{eq:attention_score}
    \end{equation}
     Intuitively, a larger $a_{\mathrm{mean}}(i)$ means several later positions place larger attention on the current position, which means an error at position \(i\) is more likely to propagate and amplify, increasing its criticality. In practice, we use \(K=5\) as an empirical setting. Rather than \(K=1\), Top-5 is less sensitive to individual outliers or attention noise. Additionally, we do not use the average attention score across all subsequent positions because it would dilute the concentrated dependencies among a few positions. Top-5 strikes a balance between robustness and sensitivity, capturing significant dependencies from later positions while reducing the influence of single extreme attention values.
\end{itemize}

A generated token is labeled as ``critical'' if it satisfies a token mismatch condition, or if either $\mathcal{H}_i$ or $a{\mathrm{mean}}(i)$ exceeds a predefined threshold. Otherwise, it is labeled as ``Non-critical''.

To mitigate the impact of the model’s own prediction errors on the critical-token judgment, we employ a \textbf{teacher-forcing evaluation} method~\cite{williams1989learning}. After assessing the importance of the token at position $i$, we replace the model’s prediction at position $i$ with the ground-truth token and continue evaluating subsequent tokens conditioned on this corrected context. This procedure suppresses error propagation effects in downstream evaluation and makes the determination of an individual token’s criticality more reflective of its actual value.

\subsubsection{Label Balance}

However, non-critical tokens constitute the majority of the dataset, leading to pronounced class imbalance. To effectively filter low-value, non-critical samples, we measure the token self-information at each generation position. Self-information, also called surprisal or information content, is a fundamental quantity in information theory that measures the information conveyed by a particular event~\cite{shannon1948mathematical}. Prior work has successfully applied it to tasks such as prompt compression by assessing the importance of lexical units~\cite{li2023compressing}. We compute the self-information of token \(x_i\) as
    \begin{equation}
    I(x_i) \;=\; -\log_2 P\bigl(x_i \mid x_0, x_1, \ldots, x_{i-1}\bigr)
    \label{eq:self_info}
    \end{equation}
where a large language model estimates the conditional probability. A larger $I(x_i)$ indicates that the token is less predictable and carries more information in the given context. Specifically, we sort all tokens labeled as non-critical by $I(x_i)$ in descending order and preferentially retain samples with higher self-information until the number of positive and negative samples reaches balance. From an information-theoretic perspective, 
this procedure selects ``hard negatives'', improving the quality of the training signal and enabling the classifier to better distinguish between critical and non-critical tokens.

\subsubsection{Classifier Training}

To accurately identify critical tokens, we designed and trained an ensemble of critical token judgers. Each judger is a 3-layer Multilayer Perceptron (MLP). The core mechanism takes the hidden state generated by a base Code-LLM for a specific token as input and outputs a binary probability indicating whether the token is critical. Let $L$ denote the number of decoder layers in the Code-LLM. For a sample $X$, we synchronously save the hidden state at position $n$ from layer $L$ of the main model and denote it by
\begin{equation}
h_n^{(L)} \;=\; \mathrm{LLM}(X)\big|_{\text{layer }L,\ \text{pos }n}.
\end{equation}
This hidden state serves as the MLP's input. The classifier produces a probability distribution over the two classes:
\begin{equation}
\mathbf{p}_n \;=\; \big[\,p_{\text{critical}},\; p_{\text{non-critical}}\,\big]^\top
\;=\; \operatorname{Classifier}\big(h_n^{(L)}\big).
\end{equation}

We optimize the classifier with the cross-entropy loss. For sample $n$ with label $y_n\in\{0,1\}$ (where $y_n=1$ indicates a \emph{critical} token), the loss is
\begin{equation}
\mathcal{L}_n \;=\; -\big( y_n\log p_{\text{critical}} + (1-y_n)\log(1-p_{\text{critical}})\big).
\end{equation}

In our experiments, we use the final-layer hidden state as the input to the classifier. Empirically, the classifier we train ranges from 4.01 to 10.01 million parameters. This is remarkably lightweight, especially when contrasted with the 1B and 13B scales of the backbone LLMs evaluated.

\begin{algorithm}[t]
\caption{Inference with Critical Token Evaluation.}
\label{alg:inference}
\noindent
\scalebox{0.92}[1]{%
\begin{minipage}{\columnwidth}
\footnotesize
\setlength{\baselineskip}{10.5pt}

\KwIn{$\textit{TargetFunctionPrompt}$, $\textit{ClassifierEnsemble}$, $\textit{Threshold } p$}
\KwOut{$\textit{FinalPrediction}$ (A complete sequence of generated code)}
\BlankLine

$\textit{Prediction} \gets [\ ]$\;
$\textit{Context} \gets \text{Retrieve}(\textit{TargetFunctionPrompt})$\;
\BlankLine

\While{not $\text{EndOfGeneration}()$}{
    $\begin{aligned}[t]
    \textit{CurrentPrompt} \gets \text{Combine}(&\textit{TargetFunctionPrompt}, \textit{Context},\\
    &\textit{Prediction})
    \end{aligned}$\;
    $\textit{current\_token}, \textit{hidden\_state} \gets \text{DecodeOneStep}(\textit{CurrentPrompt})$\;
    \BlankLine
    $\textit{is\_critical} \gets \text{CheckIfCritical}(\textit{hidden\_state}, \textit{ClassifierEnsemble}, p)$\;
    
    \If{$\textit{is\_critical}$}{
        $\begin{aligned}[t]
        \textit{Query} \gets \text{ConstructQuery}(&\textit{TargetFunctionPrompt}, \textit{Prediction},\\
        &\textit{current\_token})
        \end{aligned}$\;
        $\textit{UpdatedContext} \gets \text{Retrieve}(\textit{Query})$\;
        \BlankLine
        $\begin{aligned}[t]
        \textit{RerollPrompt} \gets \text{Combine}(&\textit{TargetFunctionPrompt}, \textit{UpdatedContext},\\
        &\textit{Prediction},\textit{current\_token})
        \end{aligned}$\;
        $\textit{current\_token} \gets \text{DecodeOneStep}(\textit{RerollPrompt})$\; \tcp{Re-decode}
        $\textit{Context} \gets \textit{UpdatedContext}$\; \tcp{Keep context}
    }
    \BlankLine
    Append $\textit{current\_token}$ to $\textit{Prediction}$\;
}
\BlankLine
\KwRet $\textit{Prediction}$\;
\end{minipage}%
}
\end{algorithm}

\subsection{Online Inference}

\subsubsection{Weighted Dense Retriever}

In repository-level code generation, we construct retrieval queries from the targeted function to obtain retrieval-augmented prompts. Typical dense retrievers~\cite{guo2022unixcoder} obtain code embedding by average-pooling token embeddings, which ignores the relative importance of token positions and thus can miss richer, generation-relevant context. Motivated by common code-generation scenarios, we propose a weighted pooling scheme whose weights follow a double-endpoint Gaussian shape.

Let a text span have length $L$ with token indices $0,1,\dots,L-1$. We define the unnormalized weight at position $i$ as the sum of two Gaussian kernels centered at the two endpoints:
\begin{equation}
\begin{aligned}
w_i &= \exp\!\Bigl(-\frac{i^{2}}{2\sigma^{2}L^{2}}\Bigr)
    +\exp\!\Bigl(-\frac{(i-(L-1))^{2}}{2\sigma^{2}L^{2}}\Bigr),\\
&\quad i=0,\dots,L-1.
\end{aligned}
\end{equation}
where $\sigma>0$ is a hyperparameter that controls the ``sharpness'' of the peaks. The first term boosts positions near the beginning (e.g., function signatures); the second term boosts positions near the end (e.g., context immediately adjacent to the ungenerated code). When $\sigma$ is relatively small, the mass concentrates at the ends; when $\sigma$ is larger, the two Gaussians overlap, and the weight profile becomes smoother or nearly flat.

To obtain a proper weighting for pooling, we apply a softmax normalization over the $w_i$:
\begin{equation}
\alpha_i \;=\; \frac{\exp(w_i)}{\displaystyle\sum_{j=0}^{L-1}\exp(w_j)}, \qquad i=0,\dots,L-1.
\end{equation}
Given token embeddings $\{\mathbf{e}_i\}_{i=0}^{L-1}$ (where $\mathbf{e}_i\in\mathbb{R}^d$), the final pooled vector is the weighted sum
\begin{equation}
\mathbf{v} \;=\; \sum_{i=0}^{L-1} \alpha_i\,\mathbf{e}_i.
\end{equation}

\noindent{This method offers two types of benefits:}
\begin{enumerate}[leftmargin=10pt]
  \item \textbf{Position-aware retrieval:} By allocating a higher weight to both the sequence head (which often contains signatures or definitions) and the sequence tail (which is near the ungenerated code), the pooled vector is more likely to retrieve context that is informative for generation. 
  \item \textbf{Low online cost:} The weights depend only on $\sigma$ and $L$, so they can be precomputed and cached for different lengths at load time, eliminating expensive computations per query and adding negligible runtime overhead.
\end{enumerate}

\subsubsection{Critical Token Guidance Inference}

To enhance the accuracy of code generation, we introduce a Critical Token-Guided Dynamic Correction strategy for inference (see Algorithm~\ref{alg:inference} for details). The core of this strategy is to perform real-time monitoring on each generated token and, upon identifying a critical token, immediately apply a correction using targeted retrieval.

The process begins with an initial retrieval using the target function prompt to obtain the initial context, which is then concatenated with the prompt to form the model's input. Subsequently, the model generates code token by token. At each step, the corresponding hidden state is extracted and fed into a pre-trained classifier ensemble for monitoring. 

If a token is classified as normal, the generation proceeds. However, if the ensemble unanimously identifies a ``Critical Token'', the system immediately initiates a dynamic correction process. It first combines the original prompt, the currently generated code, and the critical token into a new query. This query is then used to perform a second retrieval to obtain an updated, more targeted context. Finally, the model re-decodes the current step using the new context to generate a corrected token. The resulting token is appended to the generation sequence. In practice, to speed up inference, we skip the critical token check when generating line breaks and space tokens.

Notably, if a correction occurs, the updated context is retained to guide all subsequent generation steps until a new critical token appears, ensuring that the model continues to operate with the most relevant information available.

\section{Experimental Setup}

\subsection{Benchmarks}

We evaluate our approach on two repository-level benchmarks: RepoExec~\cite{le2025impacts} and CoderEval~\cite{yu2024codereval}.

\textbf{RepoExec} focuses on code generation that involves repository-level dependencies. It is designed to evaluate the ability to handle and integrate cross-file dependencies within an entire codebase while producing functionally correct code. The benchmark comprises 355 Python repository-level code generation tasks. For each task, comprehensive test suites were constructed using unit-test generation and coverage enhancement techniques.

\textbf{CoderEval} is a standardized benchmark designed for practical application scenarios. It evaluates the capability of code generation across varying levels of contextual dependency. Each task includes a function signature, a problem description, a reference solution, and a suite of unit tests; all components are executed within a unified Docker environment to ensure standardized and reproducible evaluation. In our experiments, we use the Python version, which includes 230 representative code-generation tasks drawn from real-world open-source projects.

\subsection{Baselines}

To evaluate \ours, we compare it with four representative baselines covering non-retrieval prompting, vanilla retrieval, and SOTA repository-level baselines:

\begin{itemize}[left=0pt]
    \item \textbf{RawPrompt.} A non-retrieval baseline that provides only the function signature and NL description to the generator.
    \item \textbf{RawRAG.} A vanilla \rag baseline commonly used in prior repository-level code generation studies~\cite{zhang2023repocoder,wang2025rlcoder}. It uses the RawPrompt content as retrieval query, retrieves the top-ranked context snippets, and prepends them to the prompt, representing a standard one-shot retrieval setting.
    \item \textbf{RepoCoder~\cite{zhang2023repocoder}.} This baseline is a \sota iterative \rag framework specialized for repository-level code generation. In its multi-round process, it leverages previously generated code to construct more targeted retrieval queries, thereby progressively refining the contextual information. Through this iterative context replenishment mechanism, RepoCoder enhances the model's code-generation capabilities by supplying targeted repository context.
    \item \textbf{RLCoder~\cite{wang2025rlcoder}.} This baseline introduces a retrieval optimization strategy based on Reinforcement Learning (RL). Its core mechanism enables the retriever to learn to select the most valuable context for code generation adaptively, without requiring manually labeled data. Furthermore, RLCoder incorporates a ``stop signal'' mechanism to intelligently determine which context candidates to retain, thereby striking an effective balance between retrieval efficiency and generation quality.
\end{itemize}

\subsection{Evaluation Metrics}

Consistent with prior work in the field of code generation~\cite{guo2024stop,le2025impacts,yu2024codereval}, we adopt \textbf{Pass@k} ($k=1,3,5$) as our core evaluation metric. Within our evaluation framework, a generated code sample is considered a ``correct'' solution only if it passes all associated unit tests. Based on this criterion, the Pass@k metric estimates the probability that at least one of the $k$ generated samples for a given task is correct.

\subsection{Model Selection}
In a typical RAG pipeline, there are two core components: a retriever, which selects relevant context from the repository, and a generator, which produces the final code conditioned on both the natural language description and the retrieved context. To evaluate the effectiveness and generality of our proposed method, we select representative models for both components.

\textbf{Retriever.} In our experiments, we employ UniXcoder as the retriever. As a widely adopted dense embedding model~\cite{wang2025rlcoder,deng2024r2c2}, UniXcoder~\cite{guo2022unixcoder} demonstrates excellent retrieval performance by embedding textual queries and candidate code snippets into high-dimensional dense vectors and calculating the cosine similarity between these vector embeddings. This model can efficiently retrieve the most relevant content from large-scale repositories.

\textbf{Generator.} We select two mainstream LLM families to validate the effectiveness and generalization capability of our proposed method: the DeepSeekCoder (DSCoder) series~\cite{guo2024deepseek} and the CodeLlama series~\cite{roziere2023code}. For the DSCoder series, we utilize the base version with 1.3B and 6.7B parameters. For the CodeLlama series, we utilize the Hugging Face\footnote{https://huggingface.co/} version with 7B and 13B parameters. Selecting models with varying scales and architectures enables a comprehensive evaluation of our method's applicability across different foundational generators.

\subsection{Experimental Details}

To ensure fairness and consistency in comparisons, a unified configuration is used across all experiments. In the retrieval stage, the context length is limited to 1K tokens, and the maximum number of code snippets is 10. For the training and generation stages, the temperature parameter is set as 0.6, and the maximum number of generated tokens is limited to 512. The thresholds for subsequent attention influence and uncertainty during training are configured as 0.05 and 0.8, respectively. All experiments are executed on a Linux server featuring 8 Intel(R) Xeon(R) Gold 6348 CPU cores and 8 NVIDIA A800 GPUs with 80GB of memory.

\section{Evaluation Results}

Our experiments intend to answer the following research questions (RQs):
\begin{itemize}[left=0pt]
    \item \textbf{RQ1: How does \ours compare against state-of-the-art methods in accuracy and efficiency?}
    \item \textbf{RQ2: What is the performance impact of each of our core components?}
    \item \textbf{RQ3: How sensitive is \ours to its hyperparameter settings?}
    \item \textbf{RQ4: What composition and syntactic characteristics do critical tokens exhibit?}
\end{itemize}

\subsection{RQ1: Overall Performance and Efficiency}

\begin{table*}[!t]
    \centering
    \small
    \setlength{\tabcolsep}{12pt}
    \renewcommand{\arraystretch}{1.15}
    \caption{Performance of four models on the RepoExec and CoderEval benchmarks. Superscripts in the percentages indicate the improvement of \ours over the corresponding best baseline.}

    \label{tab:main}
    \vspace{-5pt}
    \begin{tabular}{c lll lll}
    \toprule
    \multirow{2}{*}{\textbf{Method}} & \multicolumn{3}{c}{\textbf{RepoExec}} & \multicolumn{3}{c}{\textbf{CoderEval}} \\
    \cmidrule(lr){2-4} \cmidrule(lr){5-7}
    & \textbf{Pass@1} & \textbf{Pass@3} & \textbf{Pass@5} &\textbf{Pass@1} & \textbf{Pass@3} & \textbf{Pass@5} \\
    \midrule
    \RawpromptDSCOne     &7.55 &13.07 &15.77 &10.87 &17.83 &22.61 \\
    \RawragDSCOne     &8.79 &15.75 &20.00 &15.65 &23.91 &27.39 \\
    \RepocoderDSCOne  &10.03 &17.21 &21.13 &18.70 &27.39 &31.74 \\
    \RlcoderDSCOne    &10.82 &18.03 &21.97 &16.09 &26.96 &30.00 \\
    \cellcolor{blue!5}\textbf{\OursDSCOne}   
    &\cellcolor{blue!5}\textbf{11.27}\mytextsuperscript{ $ \uparrow$4.2\%}
    &\cellcolor{blue!5}\textbf{19.24}\mytextsuperscript{ $ \uparrow$6.7\%}
    &\cellcolor{blue!5}\textbf{23.10}\mytextsuperscript{ $ \uparrow$5.1\%}
    &\cellcolor{blue!5}\textbf{20.00}\mytextsuperscript{ $ \uparrow$7.0\%}
    &\cellcolor{blue!5}\textbf{28.70}\mytextsuperscript{ $ \uparrow$4.8\%}
    &\cellcolor{blue!5}\textbf{33.48}\mytextsuperscript{ $ \uparrow$5.5\%}
    \\
    \midrule
    \RawpromptDSCSeven   & 10.93 & 18.03 & 21.41 &14.35 &21.30 &26.09 \\
    \RawragDSCSeven   &14.08 &23.66 &28.73 & 20.00 & 27.83 & 31.74 \\
    \RepocoderDSCSeven&16.28 &24.76 &28.45 & 20.43 &31.74 &35.65 \\
    \RlcoderDSCSeven  &\textbf{16.68} &25.07 &28.17 & 22.61 & 31.30 &33.91 \\
    \cellcolor{blue!5}\textbf{\OursDSCSeven}    
    & \cellcolor{blue!5}\textbf{16.45\mytextsuperscript{ $ \downarrow$1.4\%}}
    & \cellcolor{blue!5}\textbf{26.20}\mytextsuperscript{ $ \uparrow$4.5\%}
    & \cellcolor{blue!5}\textbf{30.70}\mytextsuperscript{ $ \uparrow$6.9\%}
    & \cellcolor{blue!5}\textbf{23.04}\mytextsuperscript{ $ \uparrow$1.9\%}
    &\cellcolor{blue!5}\textbf{32.17}\mytextsuperscript{ $ \uparrow$1.4\%}
    & \cellcolor{blue!5}\textbf{37.83}\mytextsuperscript{ $ \uparrow$6.1\%}
    \\
    \midrule
    \RawpromptCLSeven & 9.75  & 15.38 & 18.59 & 15.65 &23.91 & 26.09 \\
    \RawragCLSeven    & 12.34 & 21.13 & 25.07 & 19.57 & 27.83 & 30.57 \\
    \RepocoderCLSeven & 13.80 & 22.11 & 27.04 & 22.61 & 29.13 & 30.43 \\
    \RlcoderCLSeven   & 13.80 & 21.41 & 25.63 & 18.26 & 24.78 & 30.87  \\
    \cellcolor{blue!5}\textbf{\OursCLSeven}
    &  \cellcolor{blue!5}\textbf{15.77}\mytextsuperscript{ $ \uparrow$14.2\%}
    & \cellcolor{blue!5}\textbf{23.94}\mytextsuperscript{ $ \uparrow$8.3\%}
    & \cellcolor{blue!5}\textbf{29.30}\mytextsuperscript{ $ \uparrow$8.4\%}
    & \cellcolor{blue!5}\textbf{23.04}\mytextsuperscript{ $ \uparrow$1.9\%}
    & \cellcolor{blue!5}\textbf{31.30}\mytextsuperscript{ $ \uparrow$7.4\%}
    & \cellcolor{blue!5}\textbf{35.65}\mytextsuperscript{ $ \uparrow$15.4\%}
    \\
    \midrule
    \RawpromptCLThirteen &9.80 &16.68 &20.00 &18.26 &25.22 &30.43 \\
    \RawragCLThirteen    &15.72 &24.60 &27.89 &24.78 &32.17 &34.78 \\
    \RepocoderCLThirteen &17.01 &27.07 &31.83 &22.61 &33.48 &35.65 \\
    \RlcoderCLThirteen   &13.75 &22.31 &26.76 &23.48 &30.43 &32.17 \\
    \cellcolor{blue!5}\textbf{\OursCLThirteen}
    & \cellcolor{blue!5}\textbf{18.59}\mytextsuperscript{ $ \uparrow$9.3\%}
    & \cellcolor{blue!5}\textbf{29.01}\mytextsuperscript{ $ \uparrow$7.2\%}
    & \cellcolor{blue!5}\textbf{32.96}\mytextsuperscript{ $ \uparrow$3.6\%}
    & \cellcolor{blue!5}\textbf{26.52}\mytextsuperscript{ $ \uparrow$7.0\%}
    & \cellcolor{blue!5}\textbf{35.65}\mytextsuperscript{ $ \uparrow$6.5\%}
    & \cellcolor{blue!5}\textbf{39.57}\mytextsuperscript{ $ \uparrow$11.0\%}
    \\
    \bottomrule
    \end{tabular}
\end{table*}

As detailed in Table~\ref{tab:main}, our framework, \ours, delivers consistent and significant performance improvements across all base models on both the RepoExec and CoderEval benchmarks. This demonstrates the broad applicability and effectiveness of our dynamic retrieval approach.

A key finding is our method's impact on multi-pass success rates. On the RepoExec dataset, for instance, while the \OursDSCSeven variant shows a marginal dip in Pass@1 score compared to the strongest baseline (\RlcoderDSCSeven), it achieves superior Pass@3 and Pass@5 results. This pattern suggests that our framework enhances the diversity of generated solutions, resulting in a higher overall success rate with increased attempts.

The value of our framework is further highlighted by the CodeLlama models. When integrated with \ours, the CodeLlama-7B model's Pass@5 score improves by a relative 8.4\% on RepoExec and a remarkable 15.4\% on CoderEval over its baseline. Similarly, the CodeLlama-13B model achieves the highest Pass@5 score in our entire evaluation (39.57\%) on CoderEval, underscoring our method's ability to unlock the full potential of powerful base models.

\begin{table}[!t]
    \centering
    \setlength{\tabcolsep}{12pt}
    \caption{Time cost analysis on CoderEval benchmark All using the DSCoder-1B model. $T_{tok}$ denotes the average inference latency per token; $T_{sam}$ represents the end-to-end execution time per sample. $N_{tok}$ measures the average number of generated tokens. $R_{gen}$ is the token-level ratio between generated code and ground truth, calculated exclusively for samples that pass all test cases.}
    \label{tab:efficiency}
    \vspace{-5pt}
    \begin{tabular}{l rrrr}
    \toprule
    \textbf{Method} & \textbf{$T_{tok}$} & \textbf{$T_{sam}$} & \textbf{$N_{tok}$} & \textbf{$R_{gen}$} \\
    \midrule
    Rawprompt & 19.3 & 2.53 & 130.82 & 83.15 \\
    Rawrag    & 19.1 & 2.21 & 108.81 & 83.38 \\
    Repocoder & 19.1 & 4.50 & 215.61 & 89.10 \\
    Rlcoder   & 19.2 & 2.28 & 106.66 & 95.38 \\
    \rowcolor{blue!5}
    \textbf{\ours} & \textbf{22.6} & \textbf{2.66} & \textbf{94.59} & \textbf{83.80} \\
    \bottomrule
    \end{tabular}
\end{table}

As illustrated in Table~\ref{tab:efficiency}, \ours achieves a superior balance between computational overhead and generation quality. We observe a slight difference in per-token latency ($T_{tok}$ = 22.6 ms, approximately 3.3 ms higher than the baselines), primarily due to the KV cache recomputation required to integrate retrieved fragments into the context. However, this operation is invoked only at sparsely identified critical token positions rather than at every generation step, keeping the associated overhead highly controllable. The selective and targeted context integration mechanism also contributes to more concise code generation. Specifically, \ours produces the fewest average tokens ($N_{tok}$ = 94.59), while achieving an $R_{gen}$ of 83.80\%, indicating that the generated code remains structurally compact and faithful to the ground truth. Compared with the iterative RepoCoder, which takes 4.50s per sample, \ours reduces end-to-end execution time to 2.66s, demonstrating that dynamic, targeted retrieval is significantly more efficient than multi-pass generation strategies.

\begin{center}
\begin{myboxc}{\textbf{RQ1 Summary:} 
\ours outperforms baselines across scales (up to \textbf{15.4\%}); Pass@5 gains suggest better accuracy and diversity. Retrieval overhead is modest because correction runs only at critical tokens.}
\end{myboxc}
\end{center}

\subsection{RQ2: Ablation Study}

\subsubsection*{Rule-level labeling}
We evaluate DSCoder-1B by omitting \emph{exactly one} labeling criterion from Section~\ref{sec:token_eval}, holding all other settings fixed. Table~\ref{tab:rule_ablation} reports Pass@$k$ relative to full \ours.

\begin{table*}[!t]
    \centering
    \setlength{\tabcolsep}{10pt}
    \renewcommand{\arraystretch}{1.1}
    \caption{Ablation on token-level labeling rules (RepoExec and CoderEval, DSCoder-1B). Each row removes one labeling criterion from Section~\ref{sec:token_eval}. Superscripts give the relative decrease versus full \ours.}
    \label{tab:rule_ablation}
    \vspace{-5pt}
    \footnotesize
    \begin{tabular}{l ccc ccc}
    \toprule
    \multirow{2}{*}{\textbf{Method}} & \multicolumn{3}{c}{\textbf{RepoExec}} & \multicolumn{3}{c}{\textbf{CoderEval}} \\
    \cmidrule(lr){2-4} \cmidrule(lr){5-7}
    & \textbf{Pass@1} & \textbf{Pass@3} & \textbf{Pass@5} &\textbf{Pass@1} & \textbf{Pass@3} & \textbf{Pass@5} \\
    \midrule
    \ours    
    &\cellcolor{blue!5}\textbf{11.27}
    &\cellcolor{blue!5}\textbf{19.24}
    &\cellcolor{blue!5}\textbf{23.10}
    &\cellcolor{blue!5}\textbf{20.00}
    &\cellcolor{blue!5}\textbf{28.70}
    &\cellcolor{blue!5}\textbf{33.48}
    \\
    \midrule
    \makecell[l]{w/o Token Mismatch}
    & 8.96\mytextsuperscript{ $ \downarrow$20.5\%}
    & 15.63\mytextsuperscript{ $ \downarrow$18.8\%}
    & 19.15\mytextsuperscript{ $ \downarrow$17.1\%}
    & 12.43\mytextsuperscript{ $ \downarrow$37.9\%}
    & 20.48\mytextsuperscript{ $ \downarrow$28.6\%}
    & 24.35\mytextsuperscript{ $ \downarrow$27.3\%}
    \\
    \makecell[l]{w/o Uncertainty}
    & 10.70\mytextsuperscript{ $ \downarrow$5.1\%}
    & 18.76\mytextsuperscript{ $ \downarrow$2.5\%}
    & 21.97\mytextsuperscript{ $ \downarrow$4.9\%}
    & 16.09\mytextsuperscript{ $ \downarrow$19.6\%}
    & 24.87\mytextsuperscript{ $ \downarrow$13.3\%}
    & 28.70\mytextsuperscript{ $ \downarrow$14.3\%}
    \\
    \makecell[l]{w/o Succ.\ Attn.\ Influence}
    & 8.51\mytextsuperscript{ $ \downarrow$24.5\%}
    & 14.99\mytextsuperscript{ $ \downarrow$22.1\%}
    & 18.03\mytextsuperscript{ $ \downarrow$22.0\%}
    & 14.09\mytextsuperscript{ $ \downarrow$29.6\%}
    & 21.83\mytextsuperscript{ $ \downarrow$23.9\%}
    & 25.65\mytextsuperscript{ $ \downarrow$23.4\%}
    \\
    \bottomrule
    \end{tabular}
\end{table*}

Table~\ref{tab:rule_ablation} supports two takeaways. First, \emph{no} single labeling criterion is redundant: dropping any one consistently lowers Pass@$k$ on both benchmarks, so mismatch, uncertainty, and subsequent-attention cues each supply non-overlapping supervision. Second, their \emph{relative} importance is benchmark-dependent rather than a fixed ranking. CoderEval behaves like a setting where errors are localized to concrete mispredictions, so mismatch-oriented supervision carries the largest margin when it is removed. RepoExec instead stresses long-horizon, cross-snippet structure, where the subsequent-attention signal shows the largest single-rule gap (on the order of low-to-mid-twenties percent relative at Pass@1). Uncertainty remains informative everywhere but is typically the least damaging omission when removed alone, which is consistent with entropy partly correlating with the other two signals.

\subsubsection*{Module-level ablation}
We next disable one runtime component at a time: dynamic critical-token inference or the position-aware weighted retriever (replaced by mean-pooled retrieval embeddings). Table~\ref{tab:component_ablation} compares each variant to full \ours.

\begin{table*}[!t]
    \centering
    \setlength{\tabcolsep}{10pt}
    \renewcommand{\arraystretch}{1.1}
    \caption{Ablation on inference and retrieval modules (RepoExec and CoderEval, DSCoder-1B). Superscripts give the relative decrease versus full \ours.}
    \label{tab:component_ablation}
    \vspace{-5pt}
    \footnotesize
    \begin{tabular}{l ccc ccc}
    \toprule
    \multirow{2}{*}{\textbf{Method}} & \multicolumn{3}{c}{\textbf{RepoExec}} & \multicolumn{3}{c}{\textbf{CoderEval}} \\
    \cmidrule(lr){2-4} \cmidrule(lr){5-7}
    & \textbf{Pass@1} & \textbf{Pass@3} & \textbf{Pass@5} &\textbf{Pass@1} & \textbf{Pass@3} & \textbf{Pass@5} \\
    \midrule
    \ours    
    &\cellcolor{blue!5}\textbf{11.27}
    &\cellcolor{blue!5}\textbf{19.24}
    &\cellcolor{blue!5}\textbf{23.10}
    &\cellcolor{blue!5}\textbf{20.00}
    &\cellcolor{blue!5}\textbf{28.70}
    &\cellcolor{blue!5}\textbf{33.48}
    \\
    \midrule 
    w/o Critical Token Inference
    & 10.42\mytextsuperscript{ $ \downarrow$7.5\%}
    & 18.31\mytextsuperscript{ $ \downarrow$5.1\%}
    & 22.25\mytextsuperscript{ $ \downarrow$3.7\%}
    & 16.96\mytextsuperscript{ $ \downarrow$15.2\%}
    & 24.78\mytextsuperscript{ $ \downarrow$15.8\%}
    & 30.00\mytextsuperscript{ $ \downarrow$10.4\%}
    \\
    \midrule 
    w/o Weighted Retriever
    & 11.04\mytextsuperscript{ $ \downarrow$2.1\%}
    & 18.28\mytextsuperscript{ $ \downarrow$5.3\%}
    & 22.54\mytextsuperscript{ $ \downarrow$2.4\%}
    & 18.26\mytextsuperscript{ $ \downarrow$8.7\%}
    & 26.52\mytextsuperscript{ $ \downarrow$7.6\%}
    & 30.43\mytextsuperscript{ $ \downarrow$9.1\%}
    \\
    \bottomrule
    \end{tabular}
\end{table*}

Table~\ref{tab:component_ablation} shows that \ours{} is neither ``retrieval-only'' nor ``inference-only'': both stages are needed. Disabling dynamic inference removes sparse, on-demand context refresh and produces the clearest quality loss, with CoderEval exhibiting substantially larger relative drops than RepoExec across Pass@$k$ (roughly mid-teens at their strongest). Reverting weighted pooling to a mean-pooled retriever yields smaller but consistent regressions, indicating that better \emph{static} context aggregation still matters even when inference is enabled. Overall, weighted retrieval sharpens what enters the prompt before and between corrections, while inference closes residual alignment gaps along the trajectory; neither substitutes fully for the other.

\begin{center}
\begin{myboxc}{\textbf{RQ2 Summary:}
Each labeling signal and each runtime stage is non-redundant: rules shift in relative priority across benchmarks, and retrieval weighting complements sparse dynamic correction rather than replacing it (Tables~\ref{tab:rule_ablation}--\ref{tab:component_ablation}).
}
\end{myboxc}
\end{center}

\begin{figure}[!t]
    \centering
    \includegraphics[width=\columnwidth]{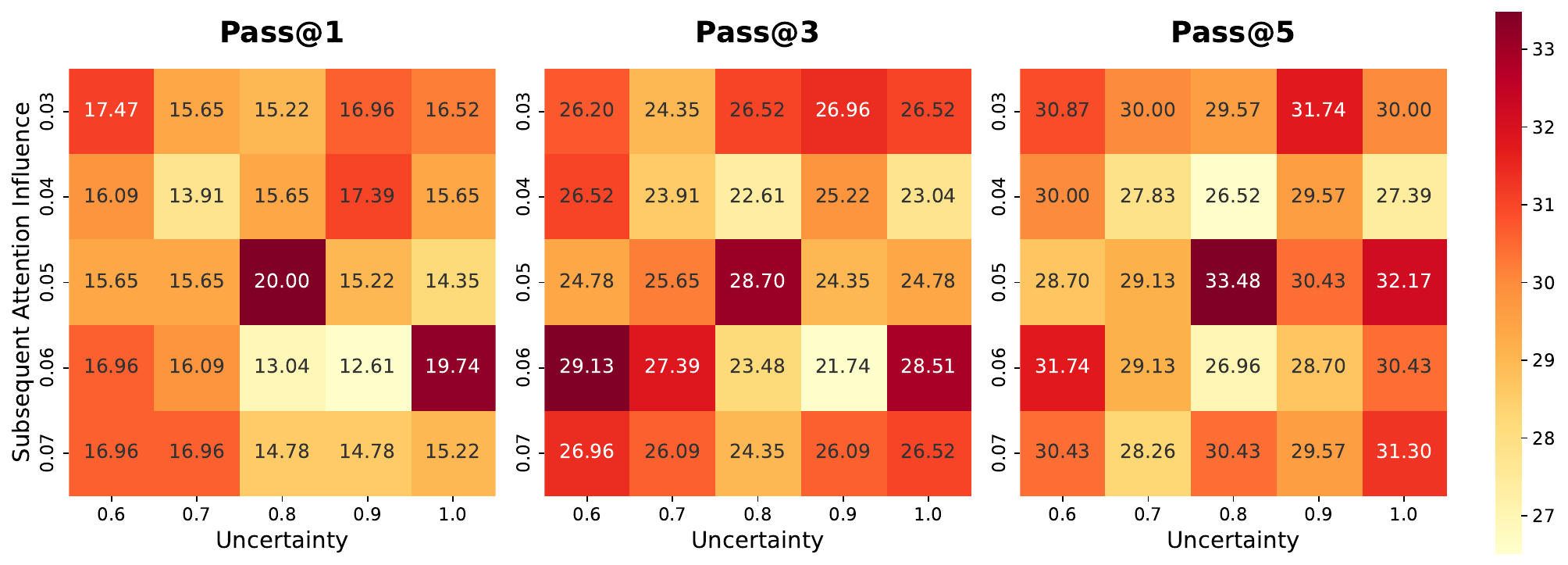}
    \vspace{-10pt}
    \caption{Performance of DSCoder-1.3B-Base model on the CoderEval benchmark under different hyperparameter configurations.}
    \label{fig:rq3}
\end{figure}

\subsection{RQ3: Sensitivity Analysis}

To explore the robustness of the \ours framework and its sensitivity to the empirical thresholds used for identifying critical tokens, we conduct a detailed sensitivity analysis on the CoderEval benchmark. As the identification of critical tokens is determined by whether the uncertainty or subsequent attention influence exceeds specific predefined thresholds, Figure~\ref{fig:rq3} presents the performance of the DeepSeek-Coder-1.3B-Base model across a variety of these hyperparameter configurations. Our analysis shows that the model achieves peak performance across the evaluation metrics when Subsequent Attention Influence is set to 0.05 and Uncertainty to 0.8. This configuration is identified as the optimal experimental setting among all evaluated candidates, yielding Pass@1 of 20.00\%, Pass@3 of 28.70\%, and Pass@5 of 33.48\%. Consequently, we adopt this specific configuration as the standard setting for all experiments in our study.

Furthermore, the performance fluctuations across the entire experimental grid are relatively marginal, which supports the claim that the directional trends of the \ours framework remain remarkably stable. Across all 25 tested configurations, the performance variance ($\approx 2.83$ for Pass@1, $\approx 3.42$ for Pass@3, and $\approx 2.65$ for Pass@5) and standard deviation ($\approx 1.68$ for Pass@1, $\approx 1.85$ for Pass@3, and $\approx 1.63$ for Pass@5) remain low, indicating that the method is insensitive to specific hyperparameter choices. This stability underscores the reliability of our dynamic, token-level context refreshing mechanism, demonstrating that \ours consistently provides the precise information the model needs, even with varying threshold settings.

\begin{center}
\begin{myboxc}{\textbf{RQ3 Summary: }\ours exhibits exceptional robustness to hyperparameter variations, maintaining stable and superior code generation performance with minimal variance across all tested threshold configurations.
}
\end{myboxc}
\end{center}

\subsection{RQ4: Composition and Syntax of Critical Tokens}

\subsubsection*{1) Critical Token Composition Analysis}

\begin{figure}[!t]
    \centering
    \includegraphics[width=0.98\columnwidth]{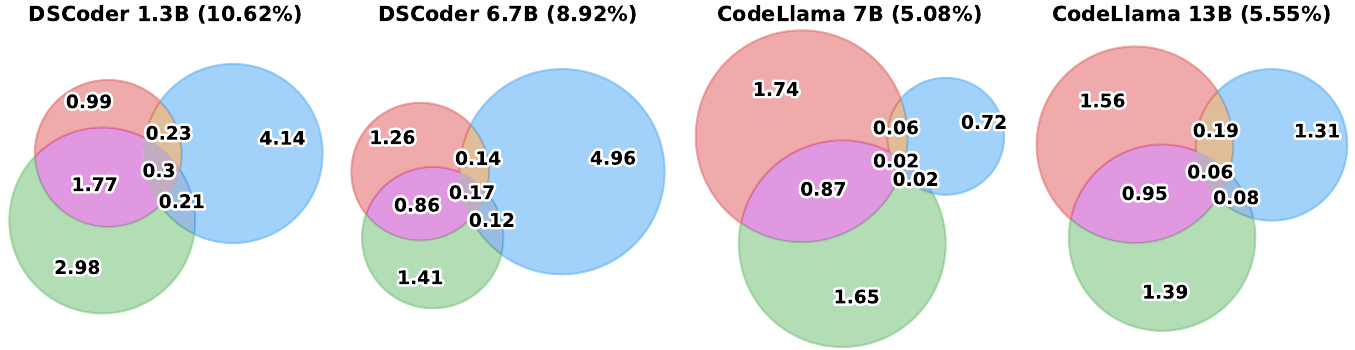}
    \caption{
        Composition and overlap of critical tokens across four models. The figure illustrates the percentage breakdown for tokens classified as
        \definecolor{mismatchred}{RGB}{229, 115, 115}
        \definecolor{attentionblue}{RGB}{100, 181, 246}
        \definecolor{uncertaintygreen}{RGB}{129, 199, 132}
        Mismatch Tokens ({\color{mismatchred}\ding{110}}),
        High Subsequent Attention Influence Tokens ({\color{attentionblue}\ding{110}}), and
        High Uncertainty Tokens({\color{uncertaintygreen}\ding{110}}).
        The percentage in each subplot's title represents the total proportion of critical tokens for that model.
    }
    \label{fig:rq4_1}
\end{figure}

We characterize the composition of critical tokens by analyzing their distribution and overlap across different models. As illustrated in Figure~\ref{fig:rq4_1}, critical tokens constitute a minimal fraction of the total, ranging from just 5.08\% to 10.62\%, with the vast majority being non-critical. A closer examination reveals a consistent overlap between mismatch tokens and high-uncertainty tokens, empirically validating that models are more prone to error when generating uncertain tokens. In contrast, the overlap between high subsequent attention influence and the other two criteria is considerably lower. This suggests that models tend to focus their attention on tokens they are confident about, rather than on those they are likely to mispredict. The proportion of tokens exhibiting all three characteristics is exceptionally low (e.g., just 0.02\% in CodeLlama 7B), indicating that models are generally efficient at avoiding deep focus on intractable tokens.

\begin{figure}[!t]
    \centering
    \includegraphics[width=1\columnwidth]{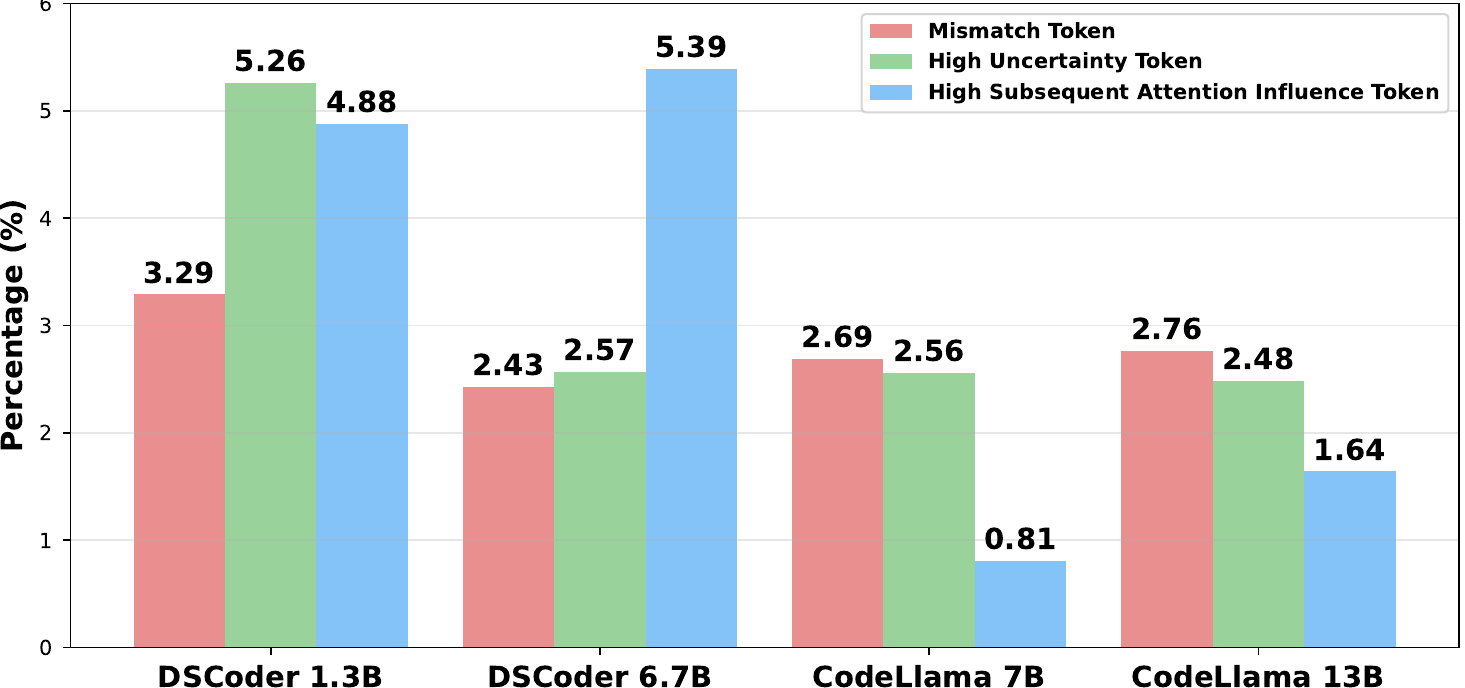}
    \caption{Comparative Analysis of Critical Token Prevalence Across Different Models}
    \label{fig:rq4_2}
\end{figure}

Furthermore, by analyzing these trends by parameter scale, Figure~\ref{fig:rq4_2} reveals that different model series exhibit distinct behaviors. In the DSCoder series, increasing the parameter count from 1B to 7B is associated with a significant decrease in both Token Mismatch (3.29\% to 2.43\%) and Uncertainty (5.26\% to 2.56\%), suggesting that larger model scale directly enhances predictive accuracy. Conversely, the CodeLlama series shows broadly stable rates for these metrics as parameter size increases from 7B to 13B, which may indicate that performance improvements have begun to plateau or are occurring elsewhere. Despite these differences, both model series share a common pattern: as parameter counts increase, the models' ``Subsequent Attention Influence'' on specific tokens tends to increase. This refined ability to identify and focus on essential information may be one of the core reasons for the improved code generation performance observed in larger-scale models.

\subsubsection*{2) Critical Token Syntactic Analysis}

\begin{figure}[t]
    \centering
    \includegraphics[width=1\columnwidth]{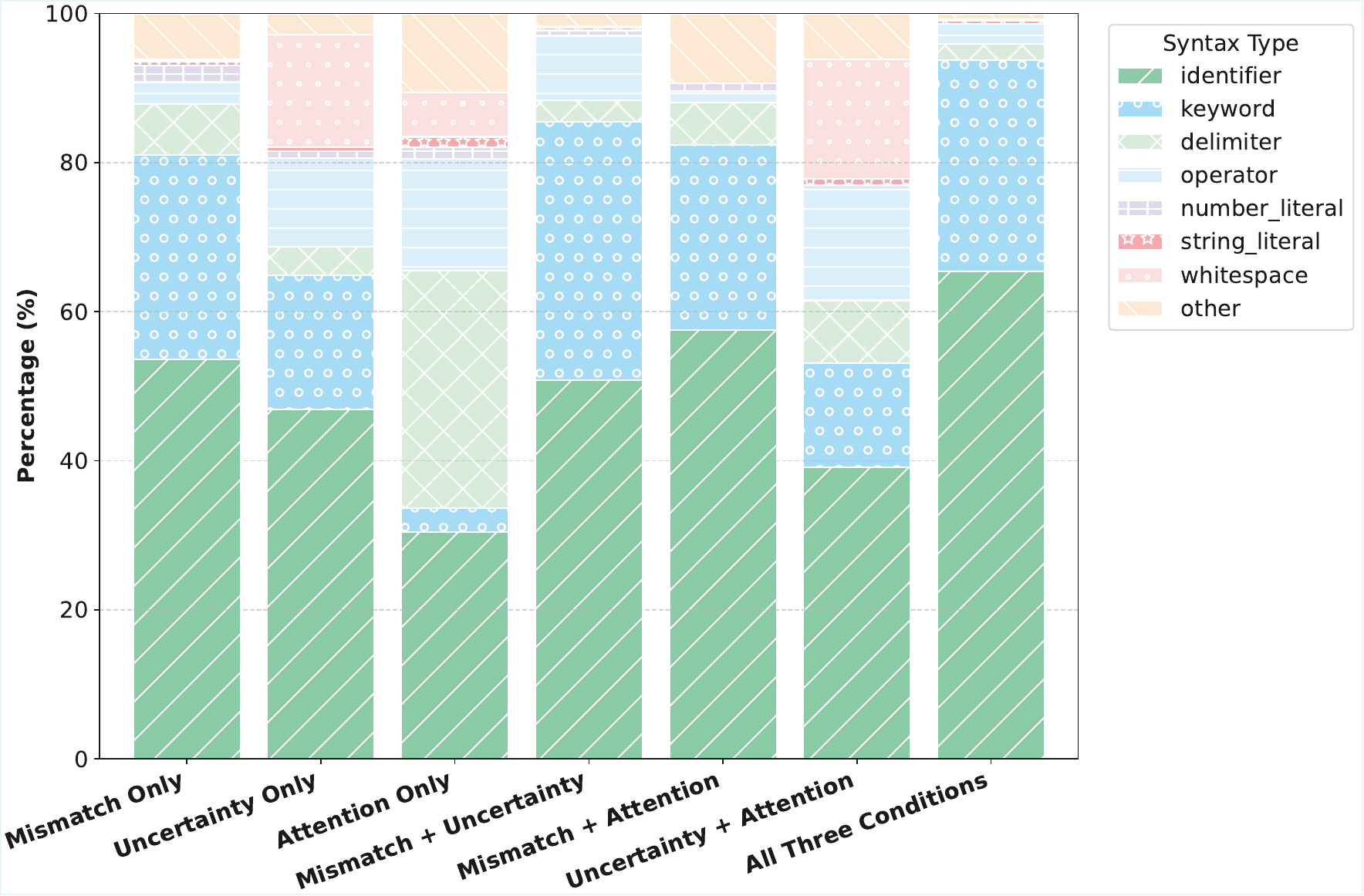}
    \vspace{-5pt}
    \caption{Absolute Percentage Distribution of Syntax Types in critical tokens, using DSCoder-1B model.}
    \label{fig:rq5_1}
\end{figure}

\begin{figure}[t]
    \centering
    \includegraphics[width=1\columnwidth]{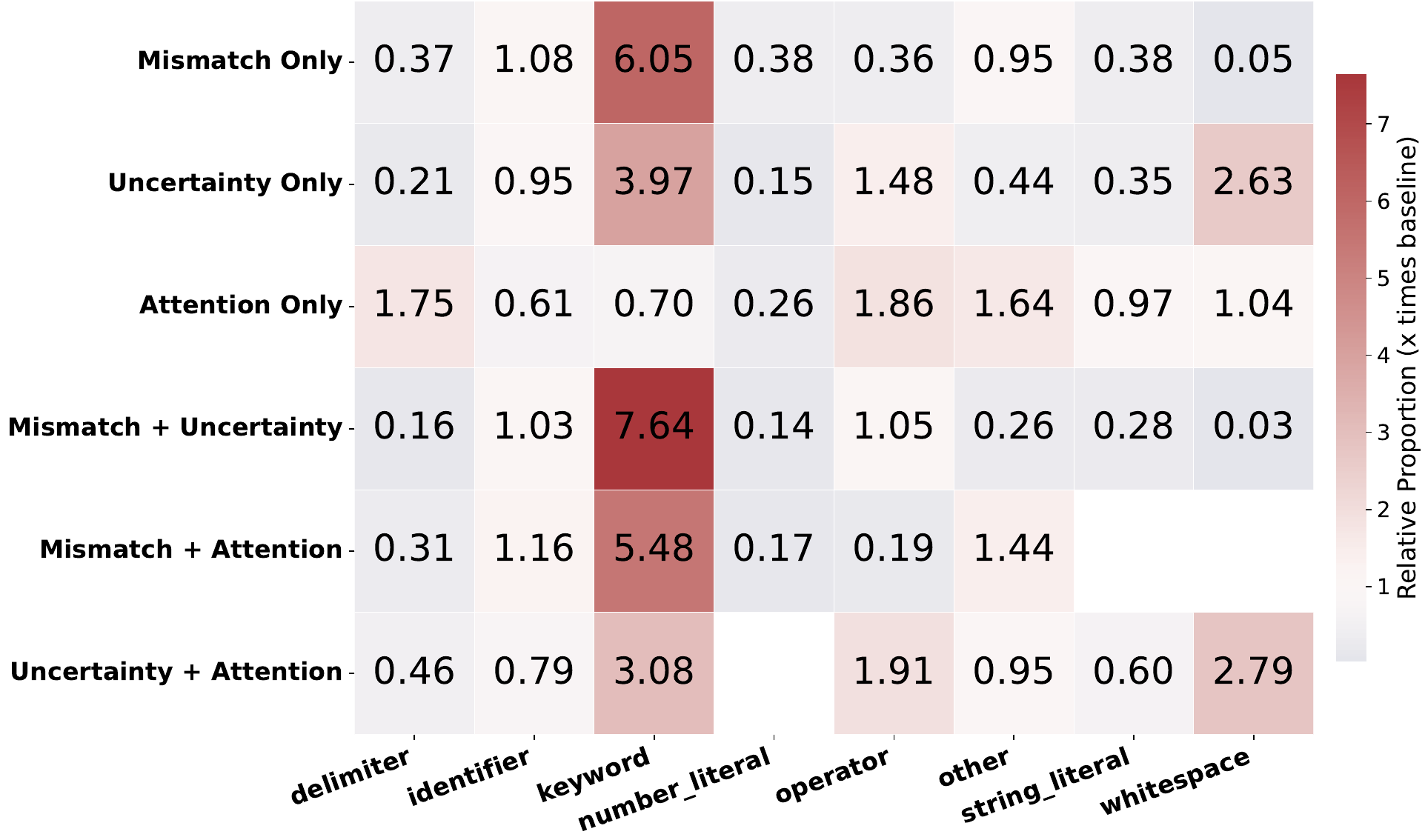}
    \caption{Relative Percentage Distribution of Syntax Types in critical tokens, using DSCoder-1B model.}
    \label{fig:rq5_2}
\end{figure}

To further explore the nature of critical tokens, we conduct a detailed analysis of their syntactic characteristics. As illustrated in Figures~\ref{fig:rq5_1} and~\ref{fig:rq5_2}, critical tokens exhibit significant syntactic diversity and a clear concentration of model attention on specific token types. For these analyses, we filter the dedent and indent syntax types.

Figure~\ref{fig:rq5_1} shows the absolute percentage distribution of syntax types within each critical token condition. A key observation is that identifiers and keywords are the predominant syntax types in most error-related conditions, such as ``Mismatch Only'' and ``Mismatch + Uncertainty''. This suggests that the model's mistakes and uncertainties most frequently occur when generating variable names and language keywords. In contrast, the ``Attention Only'' condition has a much more balanced composition, with a high proportion of delimiters and operators, indicating that the model focuses its attention not just on core logic but also on structural elements of the code.

Figure~\ref{fig:rq5_2}, which displays the relative distribution, provides deeper insights into where the models' biases lie. The heatmap highlights which syntax types are disproportionately likely to be critical compared to their normal occurrence rate. For example, keywords are exceptionally over-represented in the ``Mismatch + Uncertainty'' and ``Mismatch Only'' conditions (7.64x and 6.05x the baseline, respectively). This indicates that while identifiers are more numerous, keywords are a frequent source of combined error and uncertainty. Similarly, operators are highly over-represented in the ``Attention Only'' condition (1.86x the baseline), confirming that the model pays special attention to them.

In summary, while critical tokens are syntactically diverse, they are not random. Model errors (Mismatch) and confusion (Uncertainty) are heavily concentrated in keywords, while model focus (Attention) is disproportionately directed toward structural operators and delimiters.

\begin{center}
\begin{myboxc}{\textbf{RQ4 Summary:}
\textbf{(1) Composition.} Code-LLMs focus attention on tokens they are confident in, not on error-prone ones. As they scale, their ability to attend to essential information becomes more refined, and this improved focus, rather than fewer raw errors, drives better code generation.

\textbf{(2) Syntax.} Errors and uncertainty concentrate in critical tokens, mainly keywords, which are more likely to be mispredicted than expected. In contrast, the Code-LLM’s attention centers on structural elements such as delimiters and operators.
}
\end{myboxc}
\end{center}

\section{Case Study}

\begin{figure}[!t]
    \centering
    \includegraphics[width=\columnwidth]{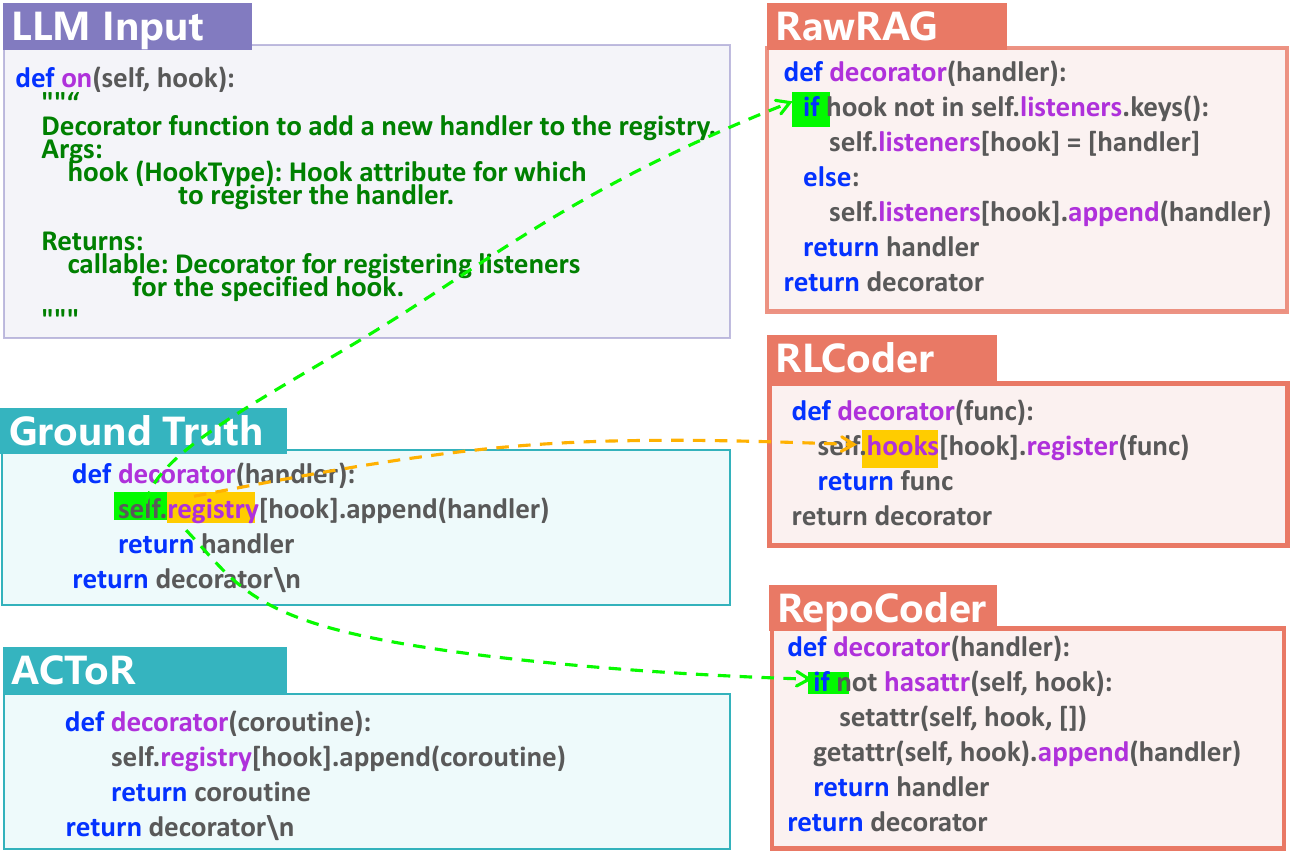}
    \vspace{-10pt}
    \caption{Case Study of Critical Token Across Different Methods}
    \label{fig:case_study}
\end{figure}

Figure~\ref{fig:case_study} illustrates a case study of various retrieval strategies in implementing a handler registration decorator, highlighting \ours\unskip's ability to generate functionally accurate code through dynamic critical token identification. When implementing the \texttt{on(self, hook)} decorator, the model must accurately identify the repository-defined attribute used for handler registration. Traditional RawRAG fails due to the limitations of static retrieval, leading to ``hallucinations'' of generic variable names like \texttt{self.listeners}. Methods such as RLCoder and RepoCoder also struggle with coarse retrieval granularity or lack dynamic adjustments, resulting in incorrect predictions, such as \texttt{self.hooks}, or overly complex and redundant logic. 

The \ours framework addresses this by identifying ``critical tokens'' during generation, triggering a precise, dynamic secondary retrieval. This on-demand mechanism allows \ours to retrieve the correct attribute, \texttt{self.registry}, in real-time. Consequently, \ours generates concise code that aligns perfectly with the Ground Truth, as shown in the comparison below, demonstrating its superiority in resolving deep repository dependencies that static methods often overlook.

\section{Threats to Validity}

\paragraph{\textbf{Internal Threats}} First, due to computational resource constraints, we evaluate a limited selection of LLMs. However, we choose representative models of varying scales, and the consistent performance gains across these models suggest that \ours\unskip's effectiveness is not model-specific. Second, to address the inherent stochasticity of LLMs, we utilize Pass@$k$ sampling strategies to mitigate the influence of sampling stochasticity. Third, whether a token is truly critical depends on its downstream impact on the final generated program, which cannot be directly observed when the token is generated. We therefore approximate critical-token labels using observable proxy signals, including token mismatch, uncertainty, and subsequent attention influence. Although these proxies may miss some functionally decisive tokens, we mitigate this threat by combining complementary signals, applying teacher forcing to reduce error-propagation effects, and validating their contribution through ablation studies.

\paragraph{\textbf{External Threats}} 

A primary concern is the computational overhead introduced by the iterative context-updating and KV cache recalculation, which may pose challenges for deployment in large-scale, real-time systems. To mitigate this, we optimize the retrieval process by avoiding redundant KV cache computations when the retrieved context remains unchanged. Our experiments demonstrate that the actual time cost remains within a controllable range for repository-level tasks. We also plan to integrate advanced KV cache optimization techniques~\cite{liu2024cachegen,10.1145/3689031.3696098,he2024let} in the future to further enhance efficiency. Another threat is the scope of benchmarks. Currently, our evaluation is focused on RepoExec and CoderEval. Although these are standard for repository-level code generation, they may not capture the full complexity of all real-world software projects. To improve generalizability, we intend to expand our evaluation to broader benchmarks in the future~\cite{guo2025omnigirl}.

\section{Related Work}

In recent years, large language models and LLM-based agents have transformed a wide range of software engineering tasks~\cite{hou2024large,fan2023large,zheng2025towards,zheng2023survey,wang2025agents,wang2025towards}. Representative applications include code generation~\cite{chen2021evaluating,chen2022codet,dehaerne2022code,zhuobigcodebench,wang2021code,liu2026shortcoderknowledgeaugmentedsyntaxoptimization,p2,p3,p4,starcoder,starcoder2,incoder}, code search~\cite{gong2025cosqa,guo2022unixcoder,wang2023you}, code summarization and comment generation~\cite{wang2025context,guo2023snippet}, and issue resolution~\cite{tao2024magis,jiang2026phoenixrepairrethinkingrepairstrategy,ye2025adverintent,guo2025omnigirl,guo2026swe,guo2026swefactoryautomatedfactoryissue,zheng2026sweprimefewertrajectoriesbetter}. Related efforts further study code review~\cite{zheng2026staticdynamicbenchmarkingrealworld}, commit message generation~\cite{shi2022race}, and repository-level code translation~\cite{wang2025repotransbench}. Building on this progress, repository-level code generation aims to produce implementations that remain consistent with an existing project rather than a standalone snippet~\cite{yu2024codereval,xie2025repost,zheng2025humanevo,wang2026arkrepobench,wang2026realsec,gu2025retrieve,jiang2025aligncoder,wang2026reporeasoner}.

\subsection{Repository-Level Code Generation}

Repository-level code generation requires models to produce code consistent with the real-world repositories~\cite{wang2025towards}. This task requires models to thoroughly understand the repository’s contents and generate accurate implementations that meet the repository’s environment and requirements. Existing work can be grouped into two main categories: retrieval-augmented methods that supply more useful context to the generator, and approaches that leverage compilation or build information for post-hoc validation and validation-driven re-retrieval.

High-quality context helps models accurately capture repository semantics, thereby improving generation quality. Prior studies focus on more effective context acquisition, exploring both retrieval-based and static-analysis-based approaches. In retrieval-based methods, RepoCoder~\cite{zhang2023repocoder} proposes an iterative pipeline that strengthens each retrieval and generation round using the previous generation. RLcoder~\cite{wang2025rlcoder} fine-tunes the retriever using feedback from the generator and introduces a stop signal strategy to discard low-relevance contexts. RepoMinCoder~\cite{li2024repomincoder} filters and reorders retrieved context based on information loss, preserving salient information while suppressing retrieval noise. In static-analysis-driven approaches, RLPG~\cite{shrivastava2023repository} manually defines ten dependency-based context sources and seven context types, and trains a classifier to identify the context requirements of tasks. Cocomic~\cite{ding2024cocomic} builds a project context graph from dependency relations and uses graph search to retrieve relevant code fragments.

Although these methods provide valuable contextual support for generation, they are generally treated as separate from retrieval, overlooking that models may require different contexts when generating different tokens. To address this, our method triggers retrieval when encountering critical tokens, providing more effective, token-specific contextual support for code generation.

\subsection{Error Analysis Of Code Generation}

Recent research has increasingly focused on understanding and mitigating errors in code generated by LLMs. Several key studies have established a foundation for this area. For instance, Wang et al.~\cite{wang2025codeerrors,2,3,4,5} conduct a thematic analysis to distill a comprehensive taxonomy of code generation errors. Their study reveals that LLMs often produce non-trivial, multi-line code-generation errors across various locations and with multiple root causes. Zhang et al.~\cite{zhang2025llm} concentrate specifically on the phenomenon of hallucinations in repository-level code generation. They define and categorize three primary hallucination categories:
task requirement conflicts, factual knowledge conflicts, and project context conflicts, which can be further divided into eight specific types. Liu et al.~\cite{liu2024exploring} likewise conduct a thematic analysis of LLM-generated code to characterize hallucinations and introduce the HalluCode benchmark for evaluating models’ ability to recognize them.

Whereas these studies focus on errors evident in final outputs, our work investigates token-level precursors to such failures. Instead of cataloguing the mistakes themselves, we analyze the underlying properties of ``critical tokens'' through the dimensions of mismatch, uncertainty, and subsequent attention influence. Our primary contribution is a fine-grained analysis that links these token-level properties to specific syntactic categories. In addition, we propose a token-level RAG framework that targets retrieval based on critical tokens to more effectively mitigate these errors.

\section{Conclusion}

In this paper, we introduce the concept of the ``critical token'', a specific point in code generation where a single error can trigger catastrophic failures. To mitigate this, we propose a novel framework that performs targeted, on-demand retrieval upon encountering a critical token, leveraging a weighted mechanism to enhance its effectiveness. In addition to this framework, we are the first to systematically quantify and analyze the properties of these tokens. Experimental results demonstrate that our approach achieves state-of-the-art performance in repository-level code generation.

\section*{Acknowledgments}
This work is supported by the National Natural Science Foundation of China (Grant No. 92582202, No. 62302534).

\bibliographystyle{IEEEtran}
\bibliography{ref}

@inproceedings{zhang2023repocoder,
  title={RepoCoder: Repository-Level Code Completion Through Iterative Retrieval and Generation},
  author={Zhang, Fengji and Chen, Bei and Zhang, Yue and Keung, Jacky and Liu, Jin and Zan, Daoguang and Mao, Yi and Lou, Jian-Guang and Chen, Weizhu},
  booktitle={Proceedings of the 2023 Conference on Empirical Methods in Natural Language Processing},
  pages={2471--2484},
  year={2023}
}

@inproceedings{wang2025rlcoder,
  title={RLCoder: Reinforcement Learning for Repository-Level Code Completion},
  author={Wang, Yanlin and Wang, Yanli and Guo, Daya and Chen, Jiachi and Zhang, Ruikai and Ma, Yuchi and Zheng, Zibin},
  booktitle={2025 IEEE/ACM 47th International Conference on Software Engineering (ICSE)},
  pages={1140--1152},
  year={2025},
  organization={IEEE}
}

@inproceedings{li2024repomincoder,
  title={Repomincoder: Improving repository-level code generation based on information loss screening},
  author={Li, Yifan and Shi, Ensheng and Zheng, Dewu and Duan, Kefeng and Chen, Jiachi and Wang, Yanlin},
  booktitle={Proceedings of the 15th Asia-Pacific Symposium on Internetware},
  pages={229--238},
  year={2024}
}

@inproceedings{zhuobigcodebench,
  title={BigCodeBench: Benchmarking Code Generation with Diverse Function Calls and Complex Instructions},
  author={Zhuo, Terry Yue and Chien, Vu Minh and Chim, Jenny and Hu, Han and Yu, Wenhao and Widyasari, Ratnadira and Yusuf, Imam Nur Bani and Zhan, Haolan and He, Junda and Paul, Indraneil and others},
  booktitle={The Thirteenth International Conference on Learning Representations}
}

@article{wang2025towards,
  title={Towards an Understanding of Context Utilization in Code Intelligence},
  author={Wang, Yanlin and Duan, Kefeng and Zheng, Dewu and Shi, Ensheng and Zhang, Fengji and Wang, Yanli and Chen, Jiachi and Liu, Xilin and Ma, Yuchi and Zhang, Hongyu and others},
  journal={arXiv preprint arXiv:2504.08734},
  year={2025}
}

@article{liu2024exploring,
  title={Exploring and evaluating hallucinations in llm-powered code generation},
  author={Liu, Fang and Liu, Yang and Shi, Lin and Huang, Houkun and Wang, Ruifeng and Yang, Zhen and Zhang, Li and Li, Zhongqi and Ma, Yuchi},
  journal={arXiv preprint arXiv:2404.00971},
  year={2024}
}

@article{zhang2025llm,
  title={Llm hallucinations in practical code generation: Phenomena, mechanism, and mitigation},
  author={Zhang, Ziyao and Wang, Chong and Wang, Yanlin and Shi, Ensheng and Ma, Yuchi and Zhong, Wanjun and Chen, Jiachi and Mao, Mingzhi and Zheng, Zibin},
  journal={Proceedings of the ACM on Software Engineering},
  volume={2},
  number={ISSTA},
  pages={481--503},
  year={2025},
  publisher={ACM New York, NY, USA}
}

@inproceedings{shrivastava2023repository,
  title={Repository-level prompt generation for large language models of code},
  author={Shrivastava, Disha and Larochelle, Hugo and Tarlow, Daniel},
  booktitle={International Conference on Machine Learning},
  pages={31693--31715},
  year={2023},
  organization={PMLR}
}

@article{chen2022codet,
  title={Codet: Code generation with generated tests},
  author={Chen, Bei and Zhang, Fengji and Nguyen, Anh and Zan, Daoguang and Lin, Zeqi and Lou, Jian-Guang and Chen, Weizhu},
  journal={arXiv preprint arXiv:2207.10397},
  year={2022}
}

@inproceedings{ding2024cocomic,
  title={CoCoMIC: Code Completion by Jointly Modeling In-file and Cross-file Context},
  author={Ding, Yangruibo and Wang, Zijian and Ahmad, Wasi U and Ramanathan, Murali Krishna and Nallapati, Ramesh and Bhatia, Parminder and Roth, Dan and Xiang, Bing},
  booktitle={Proceedings of the 2024 Joint International Conference on Computational Linguistics, Language Resources and Evaluation (LREC-COLING 2024)},
  pages={3433--3445},
  year={2024}
}

@article{xie2025repost,
  title={Repost: Scalable repository-level coding environment construction with sandbox testing},
  author={Xie, Yiqing and Xie, Alex and Sheth, Divyanshu and Liu, Pengfei and Fried, Daniel and Rose, Carolyn},
  journal={arXiv preprint arXiv:2503.07358},
  year={2025}
}

@inproceedings{le2025impacts,
  title={On the Impacts of Contexts on Repository-Level Code Generation},
  author={Le Hai, Nam and Nguyen, Dung Manh and Bui, Nghi DQ},
  booktitle={Findings of the Association for Computational Linguistics: NAACL 2025},
  pages={1496--1524},
  year={2025}
}

@inproceedings{yu2024codereval,
  title={Codereval: A benchmark of pragmatic code generation with generative pre-trained models},
  author={Yu, Hao and Shen, Bo and Ran, Dezhi and Zhang, Jiaxin and Zhang, Qi and Ma, Yuchi and Liang, Guangtai and Li, Ying and Wang, Qianxiang and Xie, Tao},
  booktitle={Proceedings of the 46th IEEE/ACM International Conference on Software Engineering},
  pages={1--12},
  year={2024}
}

@article{shannon1948mathematical,
  title={A mathematical theory of communication},
  author={Shannon, Claude E},
  journal={The Bell system technical journal},
  volume={27},
  number={3},
  pages={379--423},
  year={1948},
  publisher={Nokia Bell Labs}
}

@inproceedings{li2023compressing,
  title={Compressing Context to Enhance Inference Efficiency of Large Language Models},
  author={Li, Yucheng and Dong, Bo and Guerin, Frank and Lin, Chenghua},
  booktitle={Proceedings of the 2023 Conference on Empirical Methods in Natural Language Processing},
  pages={6342--6353},
  year={2023}
}

@article{vaswani2017attention,
  title={Attention is all you need},
  author={Vaswani, Ashish and Shazeer, Noam and Parmar, Niki and Uszkoreit, Jakob and Jones, Llion and Gomez, Aidan N and Kaiser, {\L}ukasz and Polosukhin, Illia},
  journal={Advances in neural information processing systems},
  volume={30},
  year={2017}
}

@article{williams1989learning,
  title={A learning algorithm for continually running fully recurrent neural networks},
  author={Williams, Ronald J and Zipser, David},
  journal={Neural Computation},
  volume={1},
  number={2},
  pages={270--280},
  year={1989},
  publisher={MIT Press Cambridge, MA, USA}
}

@inproceedings{guo2022unixcoder,
  title={UniXcoder: Unified Cross-Modal Pre-training for Code Representation},
  author={Guo, Daya and Lu, Shuai and Duan, Nan and Wang, Yanlin and Zhou, Ming and Yin, Jian},
  booktitle={Proceedings of the 60th Annual Meeting of the Association for Computational Linguistics (Volume 1: Long Papers)},
  pages={7212--7225},
  year={2022}
}

@inproceedings{guo2024stop,
  title={When to stop? towards efficient code generation in llms with excess token prevention},
  author={Guo, Lianghong and Wang, Yanlin and Shi, Ensheng and Zhong, Wanjun and Zhang, Hongyu and Chen, Jiachi and Zhang, Ruikai and Ma, Yuchi and Zheng, Zibin},
  booktitle={Proceedings of the 33rd ACM SIGSOFT International Symposium on Software Testing and Analysis},
  pages={1073--1085},
  year={2024}
}

@article{deng2024r2c2,
  title={R2c2-coder: Enhancing and benchmarking real-world repository-level code completion abilities of code large language models},
  author={Deng, Ken and Liu, Jiaheng and Zhu, He and Liu, Congnan and Li, Jingxin and Wang, Jiakai and Zhao, Peng and Zhang, Chenchen and Wu, Yanan and Yin, Xueqiao and others},
  journal={arXiv preprint arXiv:2406.01359},
  year={2024}
}

@article{guo2024deepseek,
  title={DeepSeek-Coder: When the Large Language Model Meets Programming--The Rise of Code Intelligence},
  author={Guo, Daya and Zhu, Qihao and Yang, Dejian and Xie, Zhenda and Dong, Kai and Zhang, Wentao and Chen, Guanting and Bi, Xiao and Wu, Yu and Li, YK and others},
  journal={arXiv preprint arXiv:2401.14196},
  year={2024}
}

@article{roziere2023code,
  title={Code llama: Open foundation models for code},
  author={Roziere, Baptiste and Gehring, Jonas and Gloeckle, Fabian and Sootla, Sten and Gat, Itai and Tan, Xiaoqing Ellen and Adi, Yossi and Liu, Jingyu and Sauvestre, Romain and Remez, Tal and others},
  journal={arXiv preprint arXiv:2308.12950},
  year={2023}
}

@inproceedings{liu2024cachegen,
  title={Cachegen: Kv cache compression and streaming for fast large language model serving},
  author={Liu, Yuhan and Li, Hanchen and Cheng, Yihua and Ray, Siddhant and Huang, Yuyang and Zhang, Qizheng and Du, Kuntai and Yao, Jiayi and Lu, Shan and Ananthanarayanan, Ganesh and others},
  booktitle={Proceedings of the ACM SIGCOMM 2024 Conference},
  pages={38--56},
  year={2024}
}

@inproceedings{10.1145/3689031.3696098,
  author = {Yao, Jiayi and Li, Hanchen and Liu, Yuhan and Ray, Siddhant and Cheng, Yihua and Zhang, Qizheng and Du, Kuntai and Lu, Shan and Jiang, Junchen},
  title = {CacheBlend: Fast Large Language Model Serving for RAG with Cached Knowledge Fusion},
  year = {2025},
  url = {https://doi.org/10.1145/3689031.3696098},
  doi = {10.1145/3689031.3696098},
  booktitle = {Proceedings of the Twentieth European Conference on Computer Systems},
  pages = {94–109},
}

@article{he2024let,
  title={Let the code llm edit itself when you edit the code},
  author={He, Zhenyu and Zhang, Jun and Luo, Shengjie and Xu, Jingjing and Zhang, Zhi and He, Di},
  journal={arXiv preprint arXiv:2407.03157},
  year={2024}
}

@inproceedings{wang2025codeerrors,
  title={Towards understanding the characteristics of code generation errors made by large language models},
  author={Wang, Zhijie and Zhou, Zijie and Song, Da and Huang, Yuheng and Chen, Shengmai and Ma, Lei and Zhang, Tianyi},
  booktitle={2025 IEEE/ACM 47th International Conference on Software Engineering (ICSE)},
  pages={717--717},
  year={2025},
  organization={IEEE Computer Society}
}

@inproceedings{fan2023large,
  title={Large language models for software engineering: Survey and open problems},
  author={Fan, Angela and Gokkaya, Beliz and Harman, Mark and Lyubarskiy, Mitya and Sengupta, Shubho and Yoo, Shin and Zhang, Jie M},
  booktitle={2023 IEEE/ACM International Conference on Software Engineering: Future of Software Engineering (ICSE-FoSE)},
  pages={31--53},
  year={2023},
  organization={IEEE}
}

@article{hou2024large,
  title={Large language models for software engineering: A systematic literature review},
  author={Hou, Xinyi and Zhao, Yanjie and Liu, Yue and Yang, Zhou and Wang, Kailong and Li, Li and Luo, Xiapu and Lo, David and Grundy, John and Wang, Haoyu},
  journal={ACM Transactions on Software Engineering and Methodology},
  volume={33},
  number={8},
  pages={1--79},
  year={2024},
  publisher={ACM New York, NY}
}

@article{chen2021evaluating,
  title={Evaluating large language models trained on code},
  author={Chen, Mark and Tworek, Jerry and Jun, Heewoo and Yuan, Qiming and Pinto, Henrique Ponde De Oliveira and Kaplan, Jared and Edwards, Harri and Burda, Yuri and Joseph, Nicholas and Brockman, Greg and others},
  journal={arXiv preprint arXiv:2107.03374},
  year={2021}
}

@article{dehaerne2022code,
  title={Code generation using machine learning: A systematic review},
  author={Dehaerne, Enrique and Dey, Bappaditya and Halder, Sandip and De Gendt, Stefan and Meert, Wannes},
  journal={Ieee Access},
  volume={10},
  pages={82434--82455},
  year={2022},
  publisher={IEEE}
}

@article{cheng2024dataflow,
  title={Dataflow-guided retrieval augmentation for repository-level code completion},
  author={Cheng, Wei and Wu, Yuhan and Hu, Wei},
  journal={arXiv preprint arXiv:2405.19782},
  year={2024}
}

@article{wu2024repoformer,
  title={Repoformer: Selective retrieval for repository-level code completion},
  author={Wu, Di and Ahmad, Wasi Uddin and Zhang, Dejiao and Ramanathan, Murali Krishna and Ma, Xiaofei},
  journal={arXiv preprint arXiv:2403.10059},
  year={2024}
}

@article{tan2024prompt,
  title={Prompt-based code completion via multi-retrieval augmented generation},
  author={Tan, Hanzhuo and Luo, Qi and Jiang, Ling and Zhan, Zizheng and Li, Jing and Zhang, Haotian and Zhang, Yuqun},
  journal={ACM Transactions on Software Engineering and Methodology},
  year={2024},
  publisher={ACM New York, NY}
}

@article{p2,
  title={Codegen: An open large language model for code with multi-turn program synthesis},
  author={Nijkamp, Erik and Pang, Bo and Hayashi, Hiroaki and Tu, Lifu and Wang, Huan and Zhou, Yingbo and Savarese, Silvio and Xiong, Caiming},
  journal={arXiv preprint arXiv:2203.13474},
  year={2022}
}

@inproceedings{p3,
  title={Intellicode compose: Code generation using transformer},
  author={Svyatkovskiy, Alexey and Deng, Shao Kun and Fu, Shengyu and Sundaresan, Neel},
  booktitle={Proceedings of the 28th ACM joint meeting on European software engineering conference and symposium on the foundations of software engineering},
  pages={1433--1443},
  year={2020}
}

@article{p4,
  title={Competition-level code generation with alphacode},
  author={Li, Yujia and Choi, David and Chung, Junyoung and Kushman, Nate and Schrittwieser, Julian and Leblond, R{\'e}mi and Eccles, Tom and Keeling, James and Gimeno, Felix and Dal Lago, Agustin and others},
  journal={Science},
  volume={378},
  number={6624},
  pages={1092--1097},
  year={2022},
  publisher={American Association for the Advancement of Science}
}

@article{starcoder,
  title={Starcoder: may the source be with you!},
  author={Li, Raymond and Allal, Loubna Ben and Zi, Yangtian and Muennighoff, Niklas and Kocetkov, Denis and Mou, Chenghao and Marone, Marc and Akiki, Christopher and Li, Jia and Chim, Jenny and others},
  journal={arXiv preprint arXiv:2305.06161},
  year={2023}
}

@article{incoder,
  title={Incoder: A generative model for code infilling and synthesis},
  author={Fried, Daniel and Aghajanyan, Armen and Lin, Jessy and Wang, Sida and Wallace, Eric and Shi, Freda and Zhong, Ruiqi and Yih, Wen-tau and Zettlemoyer, Luke and Lewis, Mike},
  journal={arXiv preprint arXiv:2204.05999},
  year={2022}
}

@article{starcoder2,
  title={Starcoder 2 and the stack v2: The next generation},
  author={Lozhkov, Anton and Li, Raymond and Allal, Loubna Ben and Cassano, Federico and Lamy-Poirier, Joel and Tazi, Nouamane and Tang, Ao and Pykhtar, Dmytro and Liu, Jiawei and Wei, Yuxiang and others},
  journal={arXiv preprint arXiv:2402.19173},
  year={2024}
}

@inproceedings{2,
  title={Idempotent code generation: Implementation, analysis, and evaluation},
  author={De Kruijf, Marc and Sankaralingam, Karthikeyan},
  booktitle={Proceedings of the 2013 IEEE/ACM International Symposium on Code Generation and Optimization (CGO)},
  pages={1--12},
  year={2013},
  organization={IEEE}
}

@ARTICLE{3,
  author={Liu, Zhijie and Tang, Yutian and Luo, Xiapu and Zhou, Yuming and Zhang, Liang Feng},
  journal={IEEE Transactions on Software Engineering}, 
  title={No Need to Lift a Finger Anymore? Assessing the Quality of Code Generation by ChatGPT}, 
  year={2024},
  volume={50},
  number={6},
  pages={1548-1584},
  doi={10.1109/TSE.2024.3392499}}

@inproceedings{4,
  title={An empirical study of code generation errors made by large language models},
  author={Song, Da and Zhou, Zijie and Wang, Zhijie and Huang, Yuheng and Chen, Shengmai and Kou, Bonan and Ma, Lei and Zhang, Tianyi},
  booktitle={7th Annual Symposium on Machine Programming},
  year={2023}
}

@article{5,
  title={Is your code generated by chatgpt really correct? rigorous evaluation of large language models for code generation},
  author={Liu, Jiawei and Xia, Chunqiu Steven and Wang, Yuyao and Zhang, Lingming},
  journal={Advances in Neural Information Processing Systems},
  volume={36},
  pages={21558--21572},
  year={2023}
}

@article{guo2026swe,
  title={SWE Data Construction, Automatically!},
  author={Guo, Lianghong and Wang, Yanlin and Li, Caihua and Tao, Wei and Yang, Pengyu and Chen, Jiachi and Song, Haoyu and Tang, Duyu and Zheng, Zibin},
  journal={Proceedings of the ACM on Software Engineering},
  volume={3},
  number={FSE},
  pages={525--546},
  year={2026},
  publisher={ACM New York, NY, USA}
}

@article{wang2026reporeasoner,
  title={RepoReasoner: Evaluating Repository-Level Code Reasoning Ability of Long-Context Language Models},
  author={Wang, Yanlin and Wang, Suiquan and Wang, Yanli and Zhang, Bowen and Guo, Daya and Chen, Jiachi and Zheng, Zibin},
  journal={Proceedings of the ACM on Software Engineering},
  volume={3},
  number={FSE},
  pages={2790--2812},
  year={2026},
  publisher={ACM New York, NY, USA}
}

@inproceedings{wang2026arkrepobench,
  title={ArkRepoBench: A Repository-Level Code Completion Benchmark for HarmonyOS Development},
  author={Wang, Yanlin and Zhang, Bowen and Wang, Yanli and Guo, Daya and Zhuo, Terry Yue and Chen, Jiachi and Liu, Mingwei and Zhang, Xingong and Zheng, Zibin},
  booktitle={Findings of the Association for Computational Linguistics: ACL 2026},
  pages={19409--19429},
  year={2026}
}

@inproceedings{wang2026realsec,
  title={RealSec-bench: A benchmark for evaluating secure code generation in real-world repositories},
  author={Wang, Yanlin and Zhang, Ziyao and Wang, Chong and Xu, Xinyi and Liu, Mingwei and Wang, Yong and Chen, Jiachi and Zheng, Zibin},
  booktitle={Findings of the Association for Computational Linguistics: ACL 2026},
  pages={35866--35883},
  year={2026}
}

@article{gu2025retrieve,
  title={What to retrieve for effective retrieval-augmented code generation? an empirical study and beyond},
  author={Gu, Wenchao and Chen, Juntao and Wang, Yanlin and Jiang, Tianyue and Li, Xingzhe and Liu, Mingwei and Liu, Xilin and Ma, Yuchi and Zheng, Zibin},
  journal={arXiv preprint arXiv:2503.20589},
  year={2025}
}

@inproceedings{jiang2025aligncoder,
  title={AlignCoder: Aligning Retrieval with Target Intent for Repository-Level Code Completion},
  author={Jiang, Tianyue and Wang, Yanli and Wang, Yanlin and Guo, Daya and Shi, Ensheng and Ma, Yuchi and Chen, Jiachi and Zheng, Zibin},
  booktitle={2025 40th IEEE/ACM International Conference on Automated Software Engineering (ASE)},
  pages={971--982},
  year={2025},
  organization={IEEE}
}

@article{gong2025cosqa,
  title={Cosqa+: Enhancing code search evaluation with a multi-choice benchmark and test-driven agents},
  author={Gong, Jing and Wu, Yanghui and Liang, Linxi and Wang, Yanlin and Chen, Jiachi and Liu, Mingwei and Zheng, Zibin},
  journal={IEEE Transactions on Software Engineering},
  volume={52},
  number={1},
  pages={206--220},
  year={2025},
  publisher={IEEE}
}

@article{wang2025repotransbench,
  title={Repotransbench: A real-world multilingual benchmark for repository-level code translation},
  author={Wang, Yanli and Wang, Yanlin and Wang, Suiquan and Guo, Daya and Chen, Jiachi and Grundy, John and Liu, Xilin and Ma, Yuchi and Mao, Mingzhi and Zhang, Hongyu and others},
  journal={IEEE Transactions on Software Engineering},
  year={2025},
  publisher={IEEE}
}

@inproceedings{zheng2025humanevo,
  title={Humanevo: An evolution-aware benchmark for more realistic evaluation of repository-level code generation},
  author={Zheng, Dewu and Wang, Yanlin and Shi, Ensheng and Zhang, Ruikai and Ma, Yuchi and Zhang, Hongyu and Zheng, Zibin},
  booktitle={2025 IEEE/ACM 47th International Conference on Software Engineering (ICSE)},
  pages={1372--1384},
  year={2025},
  organization={IEEE}
}

@article{tao2024magis,
  title={Magis: Llm-based multi-agent framework for github issue resolution},
  author={Tao, Wei and Zhou, Yucheng and Wang, Yanlin and Zhang, Wenqiang and Zhang, Hongyu and Cheng, Yu},
  journal={Advances in Neural Information Processing Systems},
  volume={37},
  pages={51963--51993},
  year={2024}
}

@article{zheng2025towards,
  title={Towards an understanding of large language models in software engineering tasks},
  author={Zheng, Zibin and Ning, Kaiwen and Zhong, Qingyuan and Chen, Jiachi and Chen, Wenqing and Guo, Lianghong and Wang, Weicheng and Wang, Yanlin},
  journal={Empirical Software Engineering},
  volume={30},
  number={2},
  pages={50},
  year={2025},
  publisher={Springer}
}

@article{zheng2023survey,
  title={A survey of large language models for code: Evolution, benchmarking, and future trends},
  author={Zheng, Zibin and Ning, Kaiwen and Wang, Yanlin and Zhang, Jingwen and Zheng, Dewu and Ye, Mingxi and Chen, Jiachi},
  journal={arXiv preprint arXiv:2311.10372},
  year={2023}
}

@inproceedings{wang2021code,
  title={Code completion by modeling flattened abstract syntax trees as graphs},
  author={Wang, Yanlin and Li, Hui},
  booktitle={Proceedings of the AAAI conference on artificial intelligence},
  volume={35},
  number={16},
  pages={14015--14023},
  year={2021}
}

@article{wang2025agents,
  title={Agents in software engineering: Survey, landscape, and vision},
  author={Wang, Yanlin and Zhong, Wanjun and Huang, Yanxian and Shi, Ensheng and Yang, Min and Chen, Jiachi and Li, Hui and Ma, Yuchi and Wang, Qianxiang and Zheng, Zibin},
  journal={Automated Software Engineering},
  volume={32},
  number={2},
  pages={70},
  year={2025},
  publisher={Springer}
}

@misc{zheng2026staticdynamicbenchmarkingrealworld,
      title={From Static to Dynamic: Benchmarking Real-World Code Review with MCR-Bench}, 
      author={Dewu Zheng and Yanlin Wang and Xiwen Wang and Kefeng Duan and Hongyu Zhang and Xilin Liu and Yuchi Ma and Zibin Zheng},
      year={2026},
      eprint={2608.27442},
      archivePrefix={arXiv},
      primaryClass={cs.SE},
      url={https://arxiv.org/abs/2608.27442}, 
}

@misc{zheng2026sweprimefewertrajectoriesbetter,
      title={SWE-Prime: Fewer Trajectories, Better Performance}, 
      author={Dewu Zheng and Ruizhe Ye and Yanlin Wang and Yang Ye and Hongyu Zhang and Ensheng Shi and Xilin Liu and Yuchi Ma and Jianxing Yu and Zibin Zheng},
      year={2026},
      eprint={2608.27449},
      archivePrefix={arXiv},
      primaryClass={cs.SE},
      url={https://arxiv.org/abs/2608.27449}, 
}

@misc{jiang2026phoenixrepairrethinkingrepairstrategy,
      title={PhoenixRepair: Rethinking Repair Strategy Exploration in Software Agents}, 
      author={Tianyue Jiang and Yanlin Wang and Xin He and Daya Guo and Jiachi Chen and Ming Wen and Ensheng Shi and Xilin Liu and Yuchi Ma and Guanbin Li},
      year={2026},
      eprint={2607.18859},
      archivePrefix={arXiv},
      primaryClass={cs.AI},
      url={https://arxiv.org/abs/2607.18859}, 
}

@misc{liu2026shortcoderknowledgeaugmentedsyntaxoptimization,
      title={ShortCoder: Knowledge-Augmented Syntax Optimization for Token-Efficient Code Generation}, 
      author={Sicong Liu and Yanxian Huang and Mingwei Liu and Jiachi Chen and Ensheng Shi and Yuchi Ma and Hongyu Zhang and Yin Zhang and Yanlin Wang},
      year={2026},
      eprint={2601.09703},
      archivePrefix={arXiv},
      primaryClass={cs.SE},
      url={https://arxiv.org/abs/2601.09703}, 
}

@article{ye2025adverintent,
  title={Adverintent-agent: Adversarial reasoning for repair based on inferred program intent},
  author={Ye, He and Yang, Aidan ZH and Hu, Chang and Wang, Yanlin and Zhang, Tao and Le Goues, Claire},
  journal={Proceedings of the ACM on Software Engineering},
  volume={2},
  number={ISSTA},
  pages={1398--1420},
  year={2025},
  publisher={ACM New York, NY, USA}
}

@article{guo2025omnigirl,
  title={Omnigirl: A multilingual and multimodal benchmark for github issue resolution},
  author={Guo, Lianghong and Tao, Wei and Jiang, Runhan and Wang, Yanlin and Chen, Jiachi and Liu, Xilin and Ma, Yuchi and Mao, Mingzhi and Zhang, Hongyu and Zheng, Zibin},
  journal={Proceedings of the ACM on Software Engineering},
  volume={2},
  number={ISSTA},
  pages={24--46},
  year={2025},
  publisher={ACM New York, NY, USA}
}

@misc{guo2026swefactoryautomatedfactoryissue,
      title={SWE-Factory: Your Automated Factory for Issue Resolution Training Data and Evaluation Benchmarks}, 
      author={Lianghong Guo and Yanlin Wang and Caihua Li and Wei Tao and Pengyu Yang and Jiachi Chen and Haoyu Song and Duyu Tang and Zibin Zheng},
      year={2026},
      eprint={2506.10954},
      archivePrefix={arXiv},
      primaryClass={cs.SE},
      url={https://arxiv.org/abs/2506.10954}, 
}

@article{wang2025context,
  title={Context-aware code summarization with multi-relational graph neural network},
  author={Wang, Yanlin and Shi, Ensheng and Du, Lun and Yang, Xiaodi and Hu, Yuxuan and Wang, Yanli and Guo, Daya and Han, Shi and Zhang, Hongyu and Zhang, Dongmei},
  journal={Automated Software Engineering},
  volume={32},
  number={1},
  pages={19},
  year={2025},
  publisher={Springer}
}

@article{guo2023snippet,
  title={Snippet comment generation based on code context expansion},
  author={Guo, Hanyang and Chen, Xiangping and Huang, Yuan and Wang, Yanlin and Ding, Xi and Zheng, Zibin and Zhou, Xiaocong and Dai, Hong-Ning},
  journal={ACM Transactions on Software Engineering and Methodology},
  volume={33},
  number={1},
  pages={1--30},
  year={2023},
  publisher={ACM New York, NY}
}

@inproceedings{wang2023you,
  title={You augment me: Exploring chatgpt-based data augmentation for semantic code search},
  author={Wang, Yanlin and Guo, Lianghong and Shi, Ensheng and Chen, Wenqing and Chen, Jiachi and Zhong, Wanjun and Wang, Menghan and Li, Hui and Zhang, Hongyu and Lyu, Ziyu and others},
  booktitle={2023 IEEE International Conference on Software Maintenance and Evolution (ICSME)},
  pages={14--25},
  year={2023},
  organization={IEEE}
}

@inproceedings{shi2022race,
  title={Race: Retrieval-augmented commit message generation},
  author={Shi, Ensheng and Wang, Yanlin and Tao, Wei and Du, Lun and Zhang, Hongyu and Han, Shi and Zhang, Dongmei and Sun, Hongbin},
  booktitle={Proceedings of the 2022 Conference on Empirical Methods in Natural Language Processing},
  pages={5520--5530},
  year={2022}
}

\end{document}